\documentclass[10pt,twocolumn]{IEEEtran}

\usepackage{graphicx}
\usepackage{epsfig} % for postscript graphics files
\usepackage{mathptmx} % assumes new font selection scheme installed
\usepackage{times} % assumes new font selection scheme installed
\usepackage{amsmath} % assumes amsmath package installed
\usepackage{amssymb}  % assumes amsmath package installed
\usepackage{color}
\usepackage[caption=false,font=footnotesize]{subfig}
\usepackage{algorithm}
\usepackage{algorithmic}
\usepackage{url}
\usepackage[mathscr]{euscript}
\DeclareMathAlphabet{\pazocal}{OMS}{zplm}{m}{n}
\DeclareMathOperator*{\argmax}{argmax}

\def\proof{\noindent{{\bf Proof: }}}

\newtheorem{theorem}{Theorem}
\newtheorem{lemma}{Lemma}

\newtheorem{definition}{Definition}

\newcommand{\E}[1]{{\mathbb{E}}\left[{#1}\right]}
\newcommand{\vc}[1]{{\mathbf{#1}}}

\newcommand{\Prob}[1]{{\mathbb{P}}\left({#1}\right)}

\newcommand{\Qv}{{\bf Q}}

\newcommand{\Zv}{{\bf Z}}

\begin{document}
\title{Dynamic Network Control for Confidential Multi-hop Communications
%\title{Throughput Optimal Scheduling with Dynamic Channel Feedback}
\thanks{Preliminary version of this paper appeared in Intl. Symposium on Modeling and Optimization in Mobile, Ad Hoc, and Wireless Networks, May
2013.} \thanks{This work was supported in part by European
Commission under Marie Curie IRSES grant PIRSES-GA-2010-269132
AGILENet.}}

\author {Yunus Sarikaya, C.~Emre Koksal,  Ozgur Ercetin %
 \thanks{Y.Sarikaya (email: sarikaya@su.sabanciuniv.edu)  and O. Ercetin (email: oercetin@sabanciuniv.edu)  are with the Department of Electronics Engineering, Faculty of Engineering and Natural Sciences, Sabanci University, 34956 Istanbul, Turkey.}\thanks{C.~E. Koksal (koksal@ece.osu.edu) is with the Department of Electrical and Computer Engineering at The Ohio State University, Columbus, OH.} }

%\author{\IEEEauthorblockN{Yunus Sarikaya, Ozgur Ercetin}
%\IEEEauthorblockA{Faculty of Natural Science and Engineering\\
%Sabanci University, Istanbul, Turkey \\
%Email: \{sarikaya,oercetin\}@sabanciuniv.edu\\} \and
%\IEEEauthorblockN{C.~Emre Koksal}
%\IEEEauthorblockA{Department of Electrical and Computer Engineering\\
%The Ohio State University, Columbus, OH\\
%Email: koksal@ece.osu.edu}}

%\author{Yunus Sarikaya, Ozgur Ercetin and C.~Emre Koksal %

%\author{\IEEEauthorblockN{Yunus Sarikaya\IEEEauthorrefmark{1}, Ozgur Ercetin\IEEEauthorrefmark{1},  C.~Emre
%Koksal\IEEEauthorrefmark{2}}
% \IEEEauthorblockA{\IEEEauthorrefmark{1}
%Fac~of Natural Science and Engineering,
%Sabanci University, Istanbul, Turkey, Email: \{sarikaya,oercetin\}@sabanciuniv.edu \\
%\IEEEauthorrefmark{2}Dept~of Electrical and Computer Engineering,
%The Ohio State University, Columbus, OH, Email: koksal@ece.osu.edu
%}}

%\author{\IEEEauthorblockN{Yunus Sarikaya\IEEEauthorrefmark{1}, Ozgur Ercetin\IEEEauthorrefmark{1},  C.~Emre
%Koksal\IEEEauthorrefmark{2}}

\maketitle

\allowdisplaybreaks \vspace{-0.3in}
\begin{abstract}
We consider the problem of resource allocation and control of
multihop networks in which multiple source-destination pairs
communicate confidential messages, to be kept confidential from the
intermediate nodes. We pose the problem as that of network utility
maximization, into which confidentiality is incorporated as an
additional quality of service constraint. We develop a simple, and
yet provably optimal dynamic control algorithm that combines flow
control, routing and end-to-end secrecy-encoding. In order to
achieve confidentiality, our scheme exploits multipath diversity and
temporal diversity due to channel variability. Our end-to-end
dynamic encoding scheme encodes confidential messages across
multiple packets, to be combined at the ultimate destination for
recovery. We first develop an optimal dynamic policy for the case in
which the number of blocks across which secrecy encoding is
performed is asymptotically large. Next, we consider encoding across
a finite number of packets, which eliminates the possibility of
achieving perfect secrecy.  For this case, we develop a dynamic
policy to choose the encoding rates for each message, based on the
instantaneous channel state information, queue states and secrecy
outage requirements. By numerical analysis, we observe that the
proposed scheme approaches the optimal rates asymptotically with
increasing block size. Finally, we address the consequences of
practical implementation issues such as infrequent queue updates and
de-centralized scheduling. We demonstrate the efficacy of our
policies by numerical studies under various network conditions.
\end{abstract}

\section{Introduction}
\label{sec:intro}

In some scenarios (e.g., tactical, financial, medical),
confidentiality of communicated information between the nodes is
necessary, so that data intended to (or originated from) a node is
not shared by any other node. Even in scenarios in which
confidentiality is not necessary, it may be dangerous to assume that
nodes will always remain uncompromised. Keeping different nodes'
information confidential can be viewed as a precaution to avoid a
captured node from gaining access to information from other
uncaptured nodes.

In this paper, we consider wireless networks in which messages are
carried between the source destination pairs cooperatively in a
multi-hop fashion via intermediate nodes. In a multihop network, as
data packets are transferred, intermediate nodes obtain all or part
of the information through directly forwarding data packets or
overhearing the transmission of nearby nodes. This poses a clear
problem when transferring confidential messages. %{\color{red}Here, we
%assume that all intermediate nodes forward data packets without
%changing its content, i.e., there are no active attacks. In this
%setting, cryptographic algorithms are used for security through
%sharing private keys, but these algorithms fail to provide perfect
%secrecy. Thus, the trend of research recently swifts to information
%theoretical secrecy, which is mathematically proven to be
%uncrackable}.
In this paper, we build efficient algorithms for confidential
multiuser communication over multihop wireless networks without the
source-destination pairs having to share any secret key a priori.
The metric we use to measure the confidentiality is the mutual
information leakage rate to the relay nodes, i.e., the {\em
equivocation rate}. We require this rate to be arbitrarily small
with high probability and impose this in the resource allocation
problem via an additional constraint.

%{\color{red}To provide secrecy in this setting, one-time pad, which
%is mathematically proven to be uncrackable, is the traditional
%solution. However, as it provides perfect secrecy, the technique is
%hard to implement practically. There are two main reasons: 1) It is
%hard to provide enough randomly-generated bits to both source and
%destination to protect all anticipated messages. 2) Most
%communication channels are noisy, which leads to errors in some bits
%and to decoding failure at the destination.}

To provide the basic intuition behind our approaches and how the
source nodes can achieve confidentiality from the relay nodes,
consider the following simple example of a diamond network given in
Fig.~\ref{fig:diamond_net}. Let the source node have a single bit of
information to be transmitted to the destination node, with {\em
perfect secrecy} (with $0$ mutual information leaked) from the relay
nodes $1$ and $2$. The issue is that, the source cannot transmit
this bit directly over one of the possible paths (through $1$ or
$2$), since either $1$ or $2$ would obtain it, violating the
confidentiality constraint. This problem can be solved by adding
random noise (i.e., randomization bit) on the information bit, and
sending the noise and the noise corrupted message over different
paths, which can then be combined at the destination. The details of
the process is as follows:

%\begin{figure}
%\begin{center}
%{\includegraphics[height=0.8in]{diamond_net}} \vspace{-0.1in}
%\caption{Diamond network} \vspace{-0.38in} \label{fig:diamond_net}
%\end{center}
%\end{figure}

\begin{figure}[t]
\centerline{
%\subfloat[Arrival and service rates]{\includegraphics[width=1.9in]{./Figures_SFC/rates_vs_V.eps}%
%\label{fig:num1}} \hfil
\subfloat[Diamond Network]{\includegraphics[height=1.3in]{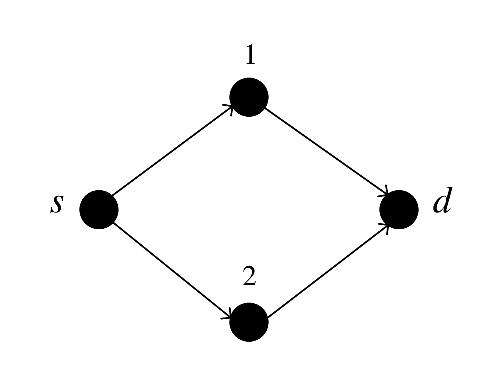}%
%subfloat[Diamond Network]{\includegraphics[height=1in]{network_models.pdf}%
\label{fig:diamond_net}} \hfil \subfloat[A multi-hop
network.]{\includegraphics[width=1.2in]{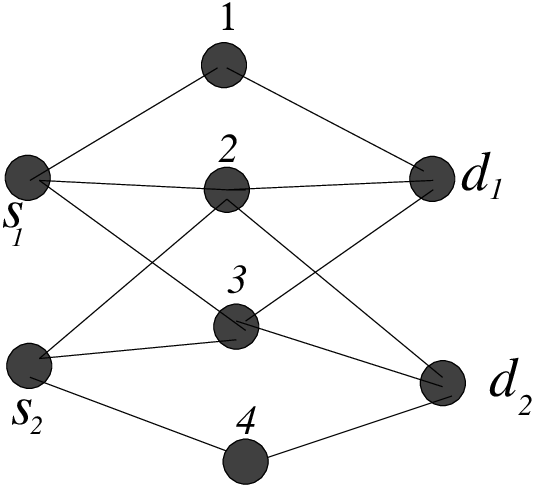}%
\label{fig:independent_set}} } \caption{Network Models }
\vspace{-0.3in}
\end{figure}

%\begin{figure}[t]
%\centerline{
%\subfloat[Arrival and service rates]{\includegraphics[width=1.9in]{./Figures_SFC/rates_vs_V.eps}%
%\label{fig:num1}} \hfil
%\subfloat[Diamond Network]{\includegraphics[height=1in]{diamond_net.eps}%
%\includegraphics[height=2in]{network_models.pdf}%
%\label{fig:diamond_net}}  \caption{Network Models } \vspace{-0.3in}
%\end{figure}

\noindent {\bf (1)} Let $b$ denote the information bit. The source generates a noise bit $N$ at random, with $\Prob{N=0}=\Prob{N=1}=\frac{1}{2}$.

\noindent {\bf (2)} Source node transmits $N$ to relay $1$ and
$b\oplus N$ to relay $2$. Then, the relay nodes forward these bits
to the destination.

\noindent {\bf (3)} Destination node reconstructs the original bit by a simple xor operation: $b=N\oplus (b\oplus N)$.

Note that with the information available to the relay nodes, there
is no way that they can make an educated guess about the information
bit, since they have \emph{zero} mutual information:
$I(b;N)=I(b;N\oplus b)=0$. Full confidentiality is achieved here at
the expense of halving of the data rate, i.e., for each information
bit the source has, the network has to carry two bits to the
destination.\footnote{It can be proved that this scheme achieves the
secrecy capacity of the diamond network, i.e., there exists no other
scheme with which one can achieve a rate higher than 1 bit/channel
use with perfect secrecy from the relay nodes. In general in a
network with $m$ parallel and independent paths, $i$th path with
capacity identical to $C_i$, the secrecy capacity is identical to
$C_s = \sum_{i=1}^m C_i- \max_{i \leq m}C_i$, hence the sum rate
minus the rate along the best relay.} Furthermore, one can see that
the existence of multiple paths from the source to the destination
is crucial to achieve perfect secrecy. However, a source cannot
route confidential information arbitrarily over the relay nodes.
Hiding information from the other nodes can be made possible by a
careful design of end-to-end coding, data routing on top of other
network mechanisms, flow control and scheduling in order for an
efficient resource utilization. Clearly, the example is highly
simplistic and ignores many important issues, which we explicitly
consider in this paper. In particular:

\noindent {\bf (a)} To achieve confidentiality, one needs to encode
blocks of information across multiple packets. We develop a novel
adaptive end-to-end encoding scheme, that takes certain observations from the
network and chooses the appropriate code rate to maintain
confidentiality for each block of data.

\noindent {\bf (b)} In a multihop network, each node possibly
overhears the transmission of a packet multiple times as it is transmitted over multiple
hops. We take into account such accumulation of information over
multiple transmissions. Thus, we need to go beyond the scenario
given in Fig. \ref{fig:diamond_net}, in which the paths are disjoint
and each intermediate node has only one path crossing.

\noindent {\bf (c)} We combine a variety of strategies developed in
the context of information theoretic secrecy with basic networking
mechanisms such as flow control and routing. Such a unifying
framework is non-existent in the literature as it pertains to
multihop information transmission. For that purpose, we model the
entire problem as that of a network utility maximization, in which
confidentiality is incorporated as an additional constraint and
develop the associated dynamic flow control, routing, and scheduling
mechanisms.

\noindent {\bf (d)} We take into account wireless channel variations
in our scheduling and routing policies as well as end-to-end
encoding scheme for confidentiality. For that purpose, we assume
that transmitters have perfect \textit{instantaneous} channel state
information (CSI) of their own channels.

\noindent {\bf Attacker model:} Each attacker is capable of tapping
into all the information transmitted and received by a single
intermediate node. Attackers are not capable of changing the content
of the information the node forwards, nor do they inject phantom
messages into the network. In our model, intermediate nodes are
entities, compliant with network operations as they properly execute
algorithms, but the messages need to be kept confidential from them.

%{\color{red} Note that even if we make use of a standard stochastic
%optimization framework, we believe that application of this
%framework in information theoretically secure multi-hop network is
%an open problem.}{\color{red}
%Note that our problem formulation differ from standard dynamic
%control formulations in two aspects: (1) Since privately encoded
%messages constitutes of randomization and private bits, we assume
%that users only gain utility from the private bits, not all the bits
%transferred to their corresponding destinations 2) We incorporate
%perfect privacy constraint into the formulation, i.e., information
%obtained by the intermediate nodes should not exceed randomization
%information encapsulated into privately encoded messages.}

We address the problem in two parts. In the first part, we ignore
the delay issue and consider the possibility of encoding across
multiple blocks of information in order to maximize the confidential
data throughput. For any given encoding rate (not necessarily
optimized for the network conditions), we provide a dynamic network
control scheme that achieves a utility, close to the maximum
achievable utility (for that particular encoding rate), subject to
{\em perfect} secrecy constraint. %, i.e., guaranteeing with
%probability $1$ that an arbitrarily low mutual information is leaked
%to the intermediate
%nodes on the confidential message. %To that end, in this paper, we
%extended our work in \cite{Sarikaya_wiopt} by considering practical
%issues due to implementation of a centralized scheduler.

The problem of network control with confidential messages has been
studied (as shall be discussed in the next section), in the past for
the single-hop setting. The main additional challenges involved in
generalizing this problem to multihop networks are dynamic
end-to-end encoding and multipath routing. Standard dynamic control
algorithms give control decisions in each time slot independently by
assuming time-scale separation, i.e., independent transmissions of
subsequent slots \cite{Georgiadis}. The confidential message is
encoded across many blocks, which implies that the time-scale
involved in physical-layer resource allocation cannot be decomposed
from the time scales involved in network-layer resource allocation,
eliminating the time-scale separation assumption of standard dynamic
control algorithms.
%Furthermore, the
%encoding rates should be determined before the transmission of the
%message without having knowledge about long-term routing decisions.
%This leads to some unique technical issues that were not addressed
%in the existing studies on network resource allocation.
A similar problem was previously addressed in \cite{Xiaojun},
where the users leave (arrive) the system, e.g., whenever their
queues empty (has one packet arrival). However, the network makes a
control decision only at the slot boundaries, delaying the network
response for at most one slot duration.  On the other hand, in our
system, due to the encoding of the confidential messages into many
blocks, the network decisions made at every slot become dependent
with each
other. This poses new and fairly different challenges in addressing the separation of time scales. Here, in order to design a cross-layer dynamic control algorithm, source needs to keep track of the rate of information obtained by each intermediate node during the transmission of encoded confidential message, and this information is required to be quantified  over each time slot independently. %, and
%transmission of the same message lasts until all packets are
%received by the destination.
In addition, the existing schemes for wireless multihop networks are
not concerned with how information ought to be spatially distributed
in the network~\cite{Y_Chen,X_Lin}. Additional ``virtual" queues
need to be maintained to keep track of the leaked information to
other nodes in the network to make sure that information from the
source node is sufficiently spatially distributed in the network.
Hence, unlike the standard multihop dynamic algorithm where the
objective is to only increase end-to-end flow rates, in our problem,
increasing the flow rate and keeping confidentiality of the messages
appear as two conflicting objectives.

%We introduce the notion of ``privacy" queues that are used to
%control information distribution along the multi-paths. In
%particular, the privacy queue method here is adapted from the
%virtual queue technique of \cite{Neely}. However, the privacy queues
%developed here are designed to spatially distribute information
%along the multi-paths, i.e., to ensure privacy constraint rather
%than average power or reliability constraints.}

In the second part, we consider practical delay requirements for
each user, which eliminates the possibility of encoding over an
arbitrarily long block. Due to finite codewords, subsequent blocks
associated with a given secrecy-encoded message cannot be decoupled.
%eliminating the possibility using standard dynamic control
%algorithms. This is due to fact that transmissions at each time slot
%is dependant on the past and future transmissions.
Also, the network mechanism cannot react to an undesirably large
rate of accumulation at a given node at a time scale faster than the
number of blocks across which the message is encoded. For the same
reason, achieving perfect secrecy for all confidential messages is
not possible. Consequently, we define the notion of \textit{secrecy
outage}, and impose a probabilistic constraint on the probability
that a message experiences a secrecy outage. We develop a dynamic
policy to choose the encoding rates for each message, based on the
instantaneous channel state information, queue states and secrecy
outage requirements. We demonstrate that our proposed scheme
approaches the maximum achievable rates asymptotically, with
increasing block size.

Finally, we investigate some practical implementation issues. In
particular, we consider the case where queue length information is
exchanged among the nodes not at every block but every $K>1$ blocks
in order to reduce the control overhead. We show that our proposed
algorithm still achieves asymptotic optimality but with longer
average queue lengths.  Another important practical limitation is
the unavailability of a centralized scheduler.  Hence, we propose a
distributed scheduling algorithm, and investigate its performance
via simulations, since the optimality can no longer be guaranteed.

%Finally, we numerically characterize the performance of dynamic
%control algorithms with respect to several network parameters.

%for a special class of interference models

%In the following, we develop our dynamic network control for
%confidential multi-hop communications by focusing on wireless
%networks. However, as will be discussed later, the proposed
%framework and developed concepts can be readily applicable to wired
%networks as well.
\section{Related Work}
\label{sec:related}

Since the pioneering work of Wyner~\cite{Wyner}, there has been a
wide interest on the variations of the wiretap channel and how to
provide secrecy in various multiuser and network information
theoretic scenarios, mainly under the single hop setting. To give a
few examples on the single hop setting, in~\cite{Gopala,Liang},
opportunistic secrecy was introduced, which allows for the
exploitation of channel variations due to fading to achieve secrecy,
even when the eavesdropper has a higher average signal-to-noise
ratio (SNR). Achieving delay-limited secrecy and outage capacity of
the wiretap channel were studied in~\cite{Gungor:infocom:10}. The
use of multiple antennas for secrecy under a variety of assumptions
has been considered in~\cite{Khisti,Shaffiee}. Multiuser
communication with secrecy using cooperative jamming and relaying in
the presence of eavesdropper was studied in~\cite{Dong}.
% The design
%of the practical codes that approach the promised capacity limits
%was investigated in~\cite{Liu}.

%Confidential multihop communication is also considered to some limited extent.
In the multihop setting,~\cite{Koyluoglu:TIT:12,Goeckel:infocom:12}
studies the secrecy capacity scaling problem. Exploitation of path
diversity in order to achieve secrecy from external eavesdroppers is
studied in~\cite{Shamir:Comm:79} and for secrecy via mobility
in~\cite{Lou:Infocom:04}. In~\cite{Cai:ISIT:02} a method is given
that modifies any given linear network code into a new code that is
secure requiring a large field size. Later,~\cite{Feldman}
generalized and simplified the method in~\cite{Cai:ISIT:02}, and
showed that the problem of making a linear network code secure is
equivalent to the problem of finding a linear code with certain
generalized distance properties. Along the same lines,~\cite{T_Cui}
investigates secure communication over wireline networks where a
node can observe one of an arbitrarily selected collection of secure
link sets. In \cite{Abuzainab}, a different notion of security
referred to as packet-level security is used, where it is sufficient
that the eavesdropper does not correctly decode the message, i.e.,
it does not guarantee full equivocation. In addition,
\cite{Abuzainab} assumes availability of private secure channels
between source and destination nodes that are not accessible to the
eavesdropper, but are more costly. The objective is to minimize the
use of private channels subject to a required level of security in
order to understand the trade-off between the security level and the
cost incurred by security.

%, and gives a linear optimization-based achievable strategies where
%random keys are canceled at intermediate non-sink nodes or injected
%at intermediate non-source node.

Despite the significant progress in information theoretic secrecy,
most of the work has focused on physical layer techniques. There has
been only a few studies on wireless network control and protocol
design with secrecy/confidentiality as a constraint on the
transmitted information. Therefore, our understanding of the
interplay between the confidentiality requirements and the critical
functionalities of wireless networks, such as scheduling, routing,
and congestion control remains limited.  There are a few number of
works on secure multi-hop communications. In \cite{Peron}, a
particular wireless relay network called the fan network is studied,
where the signal sent by a source node can be heard by all relays
via different outputs of a broadcast channel. All the relay nodes
are then connected to the destination via a perfect channel by which
destination can obtain received signal from all relays without a
delay. \cite{Infocom:Peron} considers the secret communication
between a pair of source and destination nodes in a wireless network
with authenticated relays, and derives achievable secure rates for
deterministic and Gaussian channels.

%The private message is encoded over many blocks,
%which implies that the time-scale involved in physical-layer
%resource allocation cannot be decomposed from the time scales
%involved in network-layer resource allocation, eliminating the
%time-scale separation assumption. Furthermore, the encoding rates
%should be determined before the transmission of the message without
%having knowledge about long-term routing decisions. These lead to
%some unique technical issues that were not addressed in the existing
%studies one network resource allocation.

%in~\cite{Dong,Zheng}
 Recently, in~\cite{Koksal}, we have
investigated the cross-layer resource allocation problem with
confidentiality in a cellular wireless network, where users transmit
information to the base station, confidentially from the other
users. In this paper, we consider a general multi-hop network
topology and develop dynamic network control algorithms to jointly
determine the end-to-end encoding rates, scheduling and routing.

\vspace{-0.05in}
\section{System Model} \label{sec:sys_model}

%\begin{figure}[h]
%\centering
%\begin{center}
%\end{tabular}
%\begin{tabular}{c}
% \includegraphics[width=2in]{network_model_infocom.eps}
%(b) Comparison with the passive methods
%\end{tabular}
%\vspace{-0.3in} \caption{Network model for multi-path multihop
%information transmission} \vspace{-0.2in}
%\label{fig:network_model}       % Give a unique label
%\end{center}
%\end{figure}
%as illustrated in Fig. \ref{fig:network_model}

We consider a multi-hop wireless network with $M$ source-destination
node pairs communicating with each other via intermediate relay
nodes. Let $S$ and $D$ be the sets of information ingress and egress
nodes in the network, respectively. There is no direct connection
between the nodes in $S$ and $D$, and messages from a source node to
the intended destination node are relayed by intermediate nodes in
the network. %We assume that there are no external malicious
%eavesdroppers in the system.
Let $E$ be the set of intermediate nodes which are untrusted and/or
prone to be compromised by an external attacker, i.e., $E$ denotes
the set of eavesdroppers among intermediate nodes. Note that we may
have some trusted intermediate nodes in the network. Thus, the set
of all intermediate nodes may not be the same as the set of
eavesdroppers, $E$. For ease of exposition, we consider a set of
logical links, $L$, connecting the nodes in the network, i.e., nodes
$i$ and $j$ can communicate only if link $(i,j) \in L$.

%All nodes in the network are legitimate nodes. However,
Each source node in $S$ aims to keep its information confidential
from the nodes in the set of $E$. To that end, a source node
precodes its message, divides it into multiple pieces, and sends
separate pieces over different paths to the destination. Henceforth,
none of the intermediate relay nodes in the set of $E$ will
accumulate sufficient amount of information to decode the
confidential message, even in part.

%In addition, there is a set of $L$ logical links in the network.
%Node $i$ and $j$ are connected only if link $(i,j) \in L$.

%The transmission of a privately encoded message from the source to
%the destination lasts for $N_2$ blocks, which corresponds to a total
%of $N = N_1N_2$ channel uses

We assume every channel to be i.i.d. block fading, with a block size
of $N_1$ channel uses (physical-layer symbols) where $N_1$ is
sufficiently large to allow for invoking random coding arguments
with arbitrarily low error probability. We denote the instantaneous
achievable rate of the channel between nodes $i$ and $j$ in block
$t$ by $R_{ij}(t)$, where $R_{ij}(t)$ is the maximum mutual
information between the output symbols of node $i$ and input symbols
of node $j$. We assume that the nodes are capable of obtaining
perfect instantaneous CSI. %, i.e., node $i$ has reach to the rates
%of the channels to its neighbors, $R_{ij}(t)$.
Even though our results are general for all channel state
distributions, in numerical evaluations, we use Gaussian channels,
as will be described in
Section~\ref{sec:num_results}. %Note that we assume $R_{ij}(t)$ as an
%exogenous process over which we have no control, i.e., we are not
%considering power control in this paper.
% We define the {\em instantaneous privacy
%rate} of node $i$ privately from node $j$ over block $t$ as:
%\begin{equation}
%\label{eq:privacy_rate_2} R_{ij}^p(t) = [R_i(t)-R_{ij}(t)]^+ ,
%\end{equation}
%where $[\cdot ]^+=\max(0,\cdot)$. It is shown in~\cite{fading} that
%as $N_1,N_2 \rightarrow \infty$, a long term information rate of
%$R_i^{\text{priv}}=\E{R_i^p(t)}$ is achievable by node $i$, subject
%to perfect privacy from node $j$.

%\vspace{-0.1in}
%\begin{figure}[t]
%\centering
%\begin{center}
%\end{tabular}
%\begin{tabular}{c}
% \includegraphics[width=1.2in]{multi_network.eps}
%(b) Comparison with the passive methods
%\end{tabular}
% \caption{A multi-hop
%network.}%\vspace{-0.1in}
%\label{fig:independent_set}       % Give a unique label
%\end{center}
%\end{figure}

We assume the wireless transceivers to operate in a half duplex
fashion, i.e., a node cannot transmit and receive simultaneously.
Hence, two links sharing a common ingress or egress node cannot be
active simultaneously. We define a set of links that can transmit
simultaneously as a {\em set of concurrently active links} indexed
by $v$.  Also, let $\pazocal{V}$ be the collection of all sets of
concurrently active links. Set $\pazocal{V}$ depends on the assumed
interference model. {\color{red}\footnote{We assume $ \cup_{v \in
\pazocal{V}} v$ to contain at least one path connecting any given
source-destination pair.}} For example, consider a sample multi-hop
network as shown in Fig.~\ref{fig:independent_set} with
node-exclusive interference model.
%In this example, source nodes $s_1$, $s_2$ and intermediate node 2
%can simultaneously transmit their data to nodes 1, 3, and either one
%of the destinations, respectively. In Fig.~\ref{fig:independent_set}, other
Examples of sets of concurrently active links include
$\{(s_1,1),(s_2,4),(2,d_1)\}$, $\{(s_1,1),(s_2,4),(2,d_2)\}$,
$\{(s_1,2),(1,d_1),(4,d_2)\}$, and $\{(s_2,3),(1,d_1),(4,d_2)\}$. At
the beginning of every block, the scheduler chooses a particular set
of concurrently active links. We use indicator variable ${\cal
I}_v(t)$ to represent the scheduler decision, where ${\cal I}_v(t)$
is one if set $v$ is scheduled for transmission in block $t$, and it
is zero otherwise. By definition, $\sum_{v \in \pazocal{V}} {\cal
I}_v(t) \leq 1$ for all $t>0$.

There are $M$ different flows in the network, one for each
source-destination node pair.  Each flow in the network is
identified by the index of its ingress node, and thus, the network
flow problem considered in this work can be modeled as a
multi-commodity flow problem. Let ${\cal I}^s_{ij}(t)$ be an
indicator function taking a value of 1, if link $(i,j)$ carries
commodity $s$ in block $t$. Hence, the total flow rate of commodity
$s$ over link $(i,j)$ in block $t$ is:
\begin{equation}
\mu_{ij}^s(t) = \begin{cases} R_{ij}(t), & \text { if }(i,j) \in v, {\cal I}_v(t) = 1, {\cal I}_{ij}^{s}(t)  = 1\\
0, & \text{otherwise} \end{cases}.
\end{equation}

Due to broadcast nature of wireless communications, transmissions
are overheard by unintended receivers. At every transmission,
overhearing neighboring nodes and the node which receives
information from the active link accumulate information for each
commodity $s$. Let $k$ be a node overhearing a transmission of
commodity $s$ over link $(i,j)$, i.e., there is a link $(i,k)\in L$.
Then, whenever node $k$ is not active transmitting or receiving,
i.e., no link originating or terminating at node $k$ is scheduled,
it accumulates no larger than $f_k^{s,i}(t)=\min(R_{ik}(t),\max_{j
\neq i} \mu_{ij}^s(t))$ bits of information over block $t$ over link
$(i,j)$, since overhearing information cannot exceed the actual
transmitted information. In Sections \ref{sec:wo_delay_const} and
\ref{sec:w_delay_const}, we assume that $R_{ij}(t)$ is
available causally at the centralized scheduler. In the subsequent section, we relax this assumption, and propose a distributed algorithm, i.e., nodes determine the scheduling decision by only using their local information.%for each link $(i,j)$.

%\noindent{\bf Discussion:} The perfect secrecy of multihop wireless
%communications relies on the fact that individual links experience
%statistically independent fading.  The random channel variations are
%then exploited by opportunistic algorithms such as the  control
%algorithms presented in Section \ref{sec:wo_delay_const} and
%\ref{sec:w_delay_const}.  These control algorithms can also be used
%in a wired network setting, wherein the random variations in the
%links are due to the congestion control and buffer management
%mechanisms.  The main differences in a wired network setting are:
%All links can be concurrently active, which in turn relaxes the
%scheduling constraints to complete set of links in the network; and
%the capacities of the links are constant, i.e., $R_{ij}(t) =
%R_{ij}$, for all $t$.

\section{End-to-End Confidential Encoding Rates}
\label{sec:ach_rates} %and scheduling

In this section, we describe our secrecy encoding strategy and provide an achievable confidential data rate, i.e., {\em secrecy rate} for a given source-destination pair, when the sequence of scheduling and routing decisions are given. We consider the system operation over $N_s$ blocks, which corresponds to a total of $N=N_1N_s$ channel uses. We will focus on the rate of secure data transmitted by the source node and the amount of mutual information leaked to each intermediate node by $N_s$ blocks to analyze the secrecy rate.
%Assume that, depending on the message length and routing decisions, the transmission for commodity $s$ lasts for $N_s$ blocks, i.e., the source node is done transmitting the complete message by the end of $N_s$ blocks. This corresponds to a total of $N=N_1N_s$ channel uses.

Our secrecy encoding strategy is motivated by Wyner encoding~\cite{Wyner} to provide confidentiality, which basically inserts a randomization message to the actual message to achieve equivocation. Let $C(R_s,R_s^{\textit{conf}},N)$ be a Wyner code of size $2^{NR_s}$ codewords, generated to convey a confidential message set $W_s\in \{1,\ldots , 2^{NR_s^{\textit{conf}}}\}$. In Wyner coding, a (stochastic) encoder at source node $s$ maps each confidential message $w_s \in W_s$ of size $NR_s^{\textit{conf}}$ bits to a codeword that has a length of $NR_s$ bits.
%Thus, every codeword has a length of $NR_s$ bits to convey
%$NR_s^{\textit{conf}}$ bits of confidential
%information.
Thus, we refer to $R_s^{\textit{conf}}$ as the confidential
information injection rate. Here, $N$ represents the number of
channel uses for the entire session, rather than the number of
channel uses, source $s$ is actively transmitting.\footnote{Thus,
one should view the number of channel uses $N$ in Wyner encoder
notation $C(R_s,R_s^{\textit{conf}},N)$ as a parameter that
specifies the number of bits, $NR_s^{\textit{conf}}$, at the input
and the number of bits, $NR_s$, at the  output of the encoder as
opposed to the number of channel uses that the source transmits the
message.}
%Each source node $s$ has a confidential message $W_s\in
%A decoder at the destination of source $s$ maps received sequences
%collected over $N$ channel uses, to a confidential message
%$\hat{w}_s \in W_s$. To achieve realibility, the following
%constraint on the average error probability must be met:
%\begin{align}
%    \left(P_e^s =
%    \frac{1}{2^{NR_s^{\textit{conf}}}}\sum_{w_s=1}^{2^{NR_s^{\textit{conf}}}}
%    \Prob{\hat{w}_s \neq w_s}\right) \rightarrow 0
%    \label{eq:decoding_constraint}
%\end{align}
%as $N$ goes to infinity.}

%\{1,\ldots , 2^{NR_s^{\text{conf}}}\}$ to be transmitted to an
%intended destination over $N$ channel uses.

Let the vector of symbols received by node $k$ be $\vc{Y}_k^s$. We define {\em perfect secrecy} of message $W_s$ as the following constraint to be satisfied:
\begin{equation}
\label{eq:perfect_privacy} \frac{1}{N}I(W_s,\vc{Y}_k^s) \leq
\varepsilon,% \ \ \text{as } N \rightarrow \infty.
\end{equation}
%Let us first review Wyner codes, where the main idea is to introduce
%randomness, and hence increase secrecy \cite{Wyner}. The idea relies
%on the fact that the codes take advantage of random channel errors
%(due to noise, etc.) to guarantee that information obtained by the
%eavesdropper is not adequate to decode the transmitted message.
%In order to communicate a private message $W_s$, the source
%transmits $N_1 \sum_{(s,i) \in L} \mu_{si}(t)$ bits of the
%associated binary sequence in each block $t$. Due to varying channel
%conditions and scheduling/routing decisions, source $s$ transmits
%either zero bits or $N_1 R_{si}$ bits, if $(s,i)$ link is scheduled.
%Thus, let us define $N_s$ as the number of blocks, in which source
%$s$ completes its transmission, i.e., $NR_s$ bits are transmitted in
%$N_s$ number of blocks. The total number of bits that is transmitted
%as $N_s \rightarrow \infty$ is:
for all $k \in E$ for any given $\varepsilon>0$.

%A joint routing and scheduling policy is a collection of routing
%decisions $\bigcup_{(i,j)\in L} \bigcup_{s\in S}
%\bigcup_{t=0}^{\infty} \mu_{ij}^s(t)$ which chooses the nodes that
%are actively transmitting at a given time and decides on the next
%hop for a given packet toward its destination. %Here, we assume that
%%given joint routing and scheduling policy satisfies the following
%%conditions:
In this paper, we focus on ergodic strategies, i.e., for all $(i,j)\in L$ and $s\in S$,
%This follows directly from strong law of large numbers as
\vspace{-0.1in}
\[ \lim_{N_s \rightarrow \infty} \frac{1}{N_s}
\sum_{t=1}^{N_s} \mu_{ij}^s(t) = \bar{\mu}_{ij}^s \] for some
$\bar{\mu}_{ij}^s \geq 0$, with probability 1. Hence, we also have
for all nodes $k \in E$ and $s\in S$ \vspace{-0.1in} \[ \lim_{N_s
\rightarrow \infty} \frac{1}{N_s} \sum_{t=1}^{N_s} \sum_{i \neq
k}f_k^{s,i}(t) = \bar{f}_k^s \] since $R_{ij}(t)$ is i.i.d. and
$\mu_{ij}^s(t)$ is ergodic.
%, where the expectation is over the achievable rates of all links $(s,i) \in L$.
In the following theorem, we provide the rate at which the Wyner
encoder  ought to choose the encoding parameters for a given
scheduling and routing strategy.

\begin{theorem}
   Given an ergodic joint scheduling and routing policy
% ${\cal I}_e(t),\ {\cal I}_{ij}^s(t)$ for all $t>0$ and for all $(i,j)\in L$,
confidential information injection rate achieving perfect secrecy can be lower bounded as:
       %\small
        \vspace{-0.1in}
    \begin{align}
    \label{eq:thm1}
     R_s^{conf} \geq \sum_{(s,i)\in L} \bar{\mu}_{si}^s - \max_{\forall j \in E} \ \bar{f}_j^s,
    \end{align}
    %\normalsize
    for each source-destination pair $(s,d)$, as $N_s \rightarrow \infty$.
    \label{lemma1}
\end{theorem}
The proof of this theorem is given in Appendix~\ref{proof:ach_rate}. Note that, the special case of Theorem~\ref{lemma1} that holds for deterministic channels was given in~\cite{Peron}.

Two immediate observations one can make based on Theorem~\ref{lemma1} are the following: First, Wyner encoder encapsulates secret information of $NR_s^{conf}$ bits into a block of $NR_s$, which are injected by the source into the network. Thus, we have:
\begin{equation}
\label{eq:condition_R_s} R_s \leq \lim_{N_s \rightarrow \infty}
\frac{1}{N_s}\sum_{t=1}^{N_s} \sum_{(s,i) \in L} \mu_{si}^s(t) =
\sum_{(s,i) \in L} \bar{\mu}_{si}^s .
\end{equation}
Consequently, Theorem~\ref{lemma1} implies that, for a given joint routing and scheduling strategy, rate
\[ R_s^{conf} \geq \min_{j \in E} \left\{ R_s - \bar{f}_j^{s} \right\} \]
can be achieved. Second, if there exists a node $j \in E$ through
which all possible paths between a given source $s$ and its
destination are passing, then $R_s^{conf}=0$ for that source $s$,
since $\bar{f}_j^s$ is identical to $\sum_{(s,i)\in L}
\bar{\mu}_{si}^s$ for node $j$. This underlines the necessity of the
existence of multiple
paths %\footnote{Multiple paths are those that do not share a common
%node. Note that, it is OK for disjoint paths to leak information to
%each other through links connecting pairs of nodes in those separate
%paths.}
between a source-destination pair in order to achieve a non-zero confidential data rate.

%for a given scheduling and routing policy, source $s$
%achieves a confidential data rate $R_s^{conf}=\min_{j\neq
%s}\left\{R_s-\E{f_j^{s}(t)}\right\}$.
%, where $R_s$ is the
%transmission rate of node $s$
%
%% The proof does not seem correct!!  Some implicit assumptions on routing and overhearing model??
%% Move it to appendix?
%

%Note that, $\mu_{si}(t)$ and $f_j^{s}(t)$ in Eq.~(\ref{eq:thm1}) depend on the scheduling and routing policy.

%\vspace{-0.2in}
%\begin{align}
%%NR_s = \lim_{N_1,N_s \rightarrow \infty} &\sum_{t=1}^{N_s}
%%N_1\sum_{(s,i) \in L}\mu_{si}(t)=\lim_{N_1,N_s \rightarrow \infty}
%%\frac{N}{N_s}
%%\sum_{t=1}^{N_s}\sum_{(s,i) \in L} \mu_{si}(t) \\
%R_s \geq
%%\lim_{N_1,N_s \rightarrow \infty} \frac{1}{N_s}\sum_{t=1}^{N_s}\left(
%\E{\sum_{(s,i) \in L} \mu_{si}(t)}-\delta ,
%%\right),
%\end{align}
%with probability 1 as $N_1,N_s \to \infty$ for any given $\delta
%> 0$. This follows directly from strong law of large numbers as
%\vspace{-0.1in} \[ \lim_{N_1,N_s \rightarrow \infty} \frac{1}{N_s}
%\sum_{t=1}^{N_s} \sum_{(s,i) \in L}  \mu_{si}(t) = \E{\sum_{(s,i)
%\in L}  \mu_{si}(t)} \] with probability 1, where the expectation is
%over the achievable rates of all links $(s,i) \in L$.

Before finalizing this section, there are two important notes we
would like to make. Firstly, the theorem provides a rate of encoding
for which perfect secrecy can be achieved by an appropriate choice
of secrecy encoding for a given routing and scheduling policy. It
does {\em not} imply that an end-to-end confidential information
rate of $R_s^{conf}$ is achievable. For that, it is important to
design routing and scheduling mechanisms that keeps all the queues
in the network stable and at the same time deliver all the packets
to the destination reliably. We will show how to achieve this in the
next section. Secondly, the rate provided in Theorem~\ref{lemma1} is
achieved as the number of blocks $N_s \to \infty$. This implies
that, encoding for confidentiality is done across an infinitely-long
sequence of blocks.\footnote{The number of blocks need to be large
enough for sufficient averaging of the variations in the channels.}
The network mechanisms we provide in the next section achieve
maximum achievable rate of confidential information, also over
infinitely many blocks. Following that, we incorporate a more
practical constraint of encoding over a finite number of blocks,
imposing a hard limit on the decoding delay.

%\section{Delay-Unlimited Utility Maximization}
\section{Multihop Network Control with Confidentiality}
\label{sec:wo_delay_const}
%In this section, we investigate the cases
%when a source does not have a delay requirement and encodes its very
%long messages over many blocks of information in order to maximize
%its privacy throughput. Firstly, we design a dynamic cross-layer
%algorithm for any given encoding rate which schedules the links and
%commodities with the objective of maximizing the system utility
%while controlling the information leaked to each intermediate node
%to ensure privacy of the messages. Utilizing a fixed encoding rate
%does not necessarily have to be optimal, i.e., maximizes the private
%throughput of the source. Thus, next, we propose a joint encoding
%rate selection and scheduling algorithm, which relies on encoding
%private messages after convergence of the control algorithm is
%realized.

%We first investigate the optimal scheduling and routing in \emph{delay-unlimited} case, i.e., as $N_s\rightarrow\infty$.
In the previous section, we provided the set of secrecy encoding
rates that enables confidentiality of information transmitted by the
source. However, we were not concerned with whether the packets
reach the destination or not and the mechanisms that make it happen.
In this section, our objective is to develop a stationary control
policy giving joint scheduling and routing decisions that achieves
end-to-end confidential transmission of information. To that end, we
state a network utility maximization problem and provide a scheme
that maximizes aggregate network utility while achieving perfect
secrecy over infinitely many blocks. We develop our solution based
on the following assumptions:

\begin{enumerate}
\item [A1.] We consider the large block size asymptotics, i.e., secrecy encoding is across $N_s\rightarrow\infty$ blocks.

\item [A2.] There is a centralized scheduler with the perfect knowledge of instantaneous CSI of all channels.

\item [A3.]
%messages are incoming to a source node encoded a priori at a given rate that is not necessarily optimal%, i.e., the rate may not maximize the private
%throughput of the source node.
The secrecy encoding rate is fixed: source node $s$ uses the
$C(2^{NR_s},2^{N\alpha_sR_s} ,N)$ encoder. Thus, the confidential
information injection rate is $R_s^{conf} = \alpha_sR_s$. These
rates $\{R_s^{conf},\ s\in S\}$, lie in the region of rates for
which perfect secrecy is achievable, as specified in
Theorem~\ref{lemma1}.
\end{enumerate}

Assumption A1. allows our developed mechanisms to react to an
undesirably large rate of accumulation at a given node at a time
scale faster than the number of blocks across which the message is
encoded.  Assumption A2. can be achieved by nodes sending their CSI
to the centralized scheduler at the expense of increased control
overhead.  Assumption A3. states that the a priori encoding rate of
the message may not maximize the confidential throughput of the
source node.
% Finally, Assumption A4 is needed for the developed mechanism to converge.

Next, we develop a dynamic algorithm taking as input the queue
lengths and the accumulated information at the intermediate nodes,
and gives as output the scheduled node and the admitted confidential
flows in to the queues of the sources.
%rates, respectively.
%end-to-end rates, $\{ R_s,\ s\in S\}$, and end-to-end privacy rates,
%$\{ \alpha_s R_s,\ s\in S\}$, to lie in the region of achievable
%end-to-end rates and the region of achievable end-to-end privacy
%rates, respectively.

Let $U_s(x)$ be utility obtained by source $s$ when the confidential
transmission rate is $x$ bits/channel use. We assume that
$U_s(\cdot)$ is a continuously differentiable, increasing and
strictly concave function. There is an infinite backlog at the
transport layer, which contains the secrecy-encoded messages. In
each block, source node $s$ determines the amount of encoded
information admitted to its queue at the network level. Let $A_s(t)$
be the amount of traffic injected into the queue of source $s$ at
block $t$, and $x_s=\lim_{N_s\rightarrow\infty}\frac{1}{N_s}{\sum_t
A_s(t)}$ be the long term arrival rate. Furthermore, let $Q_s(t)$
denote the queue size at ingress node $s$. Each intermediate node
keeps a separate queue $Q_i^s(t)$ for each commodity $s$.

\begin{definition}
The queues at ingress or intermediate nodes are called strongly
stable if:
\small
\begin{align*}
\limsup_{T \rightarrow \infty} \frac{1}{T} \sum_{t=0}^T \E{Q_s(t)} <
\infty \mbox{ and }\limsup_{T \rightarrow \infty} \frac{1}{T}
\sum_{t=0}^T \E{Q_i^s(t)} < \infty.
\end{align*}
\end{definition}

\normalsize Our objective is to support a traffic demand that
achieves long term confidential rates maximizing the sum of
utilities of the sources while satisfying the strong stability of
the queues.
%We assume that the utility function is continuously differentiable,
%non-decreasing and strictly concave. As mentioned before, the
%concavity assumption follows from the diminish- ing returns idea—a
%person downloading a file would feel the effect of a rate increase
%from 1 kbps to 100 kbps much more than an increase from 1 Mbps to
%1.1 Mbps although the increase is the same in both cases. It is
%immediately clear that the scena

The arrival rate, $x_s$ includes both the confidential and
randomization bits. Hence, for the arrival rate of $x_s$, the
confidential information rate is $\alpha_s x_s$, and source $s$
attains a long term expected utility of $U_s(\alpha_sx_s)$. Recall
that the rate of information obtained by any intermediate node in
the set of $E$ should not exceed the randomization rate,
$(1-\alpha_s)x_s$, to ensure perfect secrecy. Hence, the
optimization problem can be defined as follows:

\vspace{-0.2in}\small
\begin{align}
\label{objective_func_fixed}
    &\max_{A_s(t), {\cal I}_v(t), {\cal I}_{ij}^s(t)} \sum_{s\in S} U_s(\alpha_s x_s) \\
    \label{stability_const_fixed}
    &\,\,\,\text{s.t.  } x_s \leq \sum_{\{i|(s,i)\in L\}} \bar{\mu}_{si}^s,
    \,    \forall s \in S\\
    \label{flow_conservation_const_fixed}
    &\,\, \sum_{\{j|(i,j)\in L\}} \bar{\mu}_{ij}^s - \sum_{\{i|(j,i)\in L\}} \bar{\mu}_{ji}^s \geq 0, \, \forall \ i \notin S,D \\
    \label{secrecy_const_fixed}
    &\,\, \bar{f}_j^{s} \leq (1-\alpha_s) x_s, \, \forall s \in S, \forall j \in E ,
\end{align}
\normalsize

Constraint~\eqref{stability_const_fixed} ensures the stability of
the queues at the source nodes;
Constraint~\eqref{flow_conservation_const_fixed} is the flow
conservation constraint at the intermediate nodes%, i.e., the rate of
%incoming packets at intermediate node should be smaller that the
%rate of outgoing packets from that node
 ; and
Constraint~\eqref{secrecy_const_fixed} is the confidentiality
constraint, %which ensures that the information obtained by any of
%the intermediate nodes does not exceed the rate of the
%randomization message $(1-\alpha_s)x_s$,
implying that the requirement in~\eqref{eq:perfect_privacy} is met. %If one chooses the utility functions to be linear, the solution of the problem gives us any weighted sum of the private rates achieved by the sources.

To solve the optimization problem
\eqref{objective_func_fixed}-\eqref{secrecy_const_fixed}, we employ
a cross-layer dynamic control algorithm based on the stochastic
network optimization framework developed in \cite{Georgiadis}. %, since the
%objective function in \eqref{objective_func_fixed} is a concave
%function
This framework allows the solution of a long-term stochastic
optimization problem without requiring the explicit characterization
of the achievable rate regions.\footnote{Note that, while we know
that the arrival rates lie in the region of rates for perfect
secrecy (Theorem~\ref{lemma1}), we do not know the achievable
end-to-end rates with confidentiality.}
%, {\color{red}which requires the
%long-term averages of the scheduling decisions}.

The queue evolution equation for each queue can be stated as
follows:

\vspace{-0.2in} \small
\begin{align}
    Q_s(t+1) &= \left[ Q_s(t) - \sum_{\{i|(s,i)\in L\}}
    \mu_{si}^s(t)
    \right]^+ + A_s(t), \nonumber \\
    Q_i^s(t+1) &= \left[ Q_i^s(t) - \sum_{\{j|(i,j)\in L\}} \mu_{ij}^s(t)\right]^+ +  \sum_{\{j|(j,i)\in L\}}
    \mu_{ji}^s(t), \ \forall i \neq S,D, \nonumber%\label{eq:actual_queues}
\end{align}
\normalsize  where $[.]^+$ denotes the projection of the term to
$[0,+\infty)$. To meet Constraint~\eqref{secrecy_const_fixed}, we
maintain a virtual queue:

\vspace{-0.1in} \small
\begin{align}
    Z_j^s(t+1) &= \left[ Z_j^s(t) + \sum_{i \neq j}f_j^{s,i}(t) - (1-\alpha_s)A_s(t)
    \right]^+. \label{update_information}
\end{align}
\normalsize  Strong stability of this queue ensures that the
constraint is satisfied~\cite{Georgiadis}, i.e., perfect secrecy is
achieved in our case. Note that to perform the update in
\eqref{update_information}, nodes need to have access to
instantaneous CSI of all neighboring nodes.

\vspace{-0.0in}  \noindent {\bf Control Algorithm 1 (with Perfect
CSI):} The algorithm executes the following steps in each block $t$:
\begin{enumerate}
\item[\bf (1)] {\bf Flow control:} For some $H>0$, each
source $s$ injects $A_s(t)$ bits into its queues, where
\vspace{-0.1in} \small
\begin{align*}
A_s(t)=  \argmax_{A}\left\{HU_s(\alpha_s A)- Q_s(t)A + \sum_{j\in E}
Z_j^s(t) (1-\alpha_s)A \right\}.
\end{align*}

\normalsize
\item[\bf (2)] {\bf Scheduling:} In each block, $t$, the scheduler
chooses the set of links $v$ if ${\cal I}_v(t) = 1$ and flow $s$ on
the link $(i,j) \in v$ if ${\cal I }_{ij}^s(t) = 1 $, where

\vspace{-0.1in} \footnotesize
\begin{align*}
%&\left({\cal I}_e(t), {\cal I}_{ij}^s(t)\right) = 1\ \forall (i,j)\in e\ \text{and $s$, where} \\
&(s,v)=\argmax_{s\in S,v\in \pazocal{V}} \left\{ \sum_{(i,j)\in
v}(Q_i^s(t)-Q_j^s(t))\mu_{ij}^s(t) - \sum_{j \in E} \sum_{i \neq j}
Z_j^s(t)f_j^{s,i}(t) \right\}.
\end{align*}
\normalsize
 \vspace{0.0in}

\end{enumerate}
\normalsize

%%%%
% Burada hatalar var!
%%%%
\vspace{-0.1in} \noindent {\bf Optimality of Control Algorithm:}
Now, we show that our proposed dynamic
control algorithm can achieve a performance arbitrarily close to the
optimal solution while keeping the queue backlogs bounded.

\begin{theorem}
\label{thm:optimalcontrol-1}
 If $R_{ij}(t)<\infty$ for all $(i,j)$ links and for all $t$ blocks, then control
algorithm satisfies:

\small \vspace{-0.2in}
\begin{align*}
\liminf_{t\rightarrow\infty}\frac{1}{t}\sum_{\tau=0}^{t-1}\sum_{s\in
S}
\E{U_s(\tau)} &\geqslant U^* - \frac{B}{H} \\
\limsup_{t\rightarrow\infty}\frac{1}{t}\sum_{\tau=0}^{t-1}\sum_{s\in
S}
\E{Q_s(\tau)} &\leqslant \frac{B+H(\bar{U}-U^*)}{\epsilon_1} \\
\limsup_{t\rightarrow\infty}\frac{1}{t}\sum_{\tau=0}^{t-1}\sum_{s\in
S}\sum_{i \notin S,D}
\E{Q_i^s(\tau)} &\leqslant \frac{B+H(\bar{U}-U^*)}{\epsilon_2} \\
\end{align*}

\normalsize\vspace{-0.2in} \noindent where $B,\epsilon_1,\epsilon_2$
are positive constants, $U^*$ is the optimal aggregate utility,
i.e., the solution of
(\ref{objective_func_fixed}-\ref{secrecy_const_fixed}), and
$\bar{U}$ is the maximum possible instantaneous aggregate utility.
%Theorem \ref{thm:optimalcontrol-1} shows that the algorithm results
%in a time average penalty that is within $O(1/H)$ of optimality,
%with a corresponding $O(H)$ tradeoff in average queue size. %That is
%to say, with the choice of large $H$, the algorithm gets closer to
%the optimal aggregate utility at the expense of larger average queue
%sizes.
\end{theorem}

%The proof of Theorem \ref{thm:optimalcontrol-1} is given in
%Appendix~\ref{proof:optimalcontrol-1}.
%Due to space limitations, here we omit the proof, which can be found
%in our technical report~\cite{Tech_report}.
The proof of Theorem \ref{thm:optimalcontrol-1} is given in Appendix
\ref{proof:optimalcontrol-1}. This theorem shows that it is possible
to get arbitrarily close to the optimal utility by choosing $H$
sufficiently large at the expense of proportionally increased
average queue sizes. However, since all queues remain bounded, the destination is receiving packets at the rate as they are injected at the source as $N_s\to \infty$.
%Arbitrarily close to the optimal utility implies that any rate in the achievable rate region is achieved as long as it lies within the secrecy rate region specified by Theorem 1.

%We acknowledge that here the problem is investigated in
%a simplistic setting considering the assumptions presented in the
%beginning of this section. In the subsequent sections, we relax
%those assumptions and build more realistic algorithms. The scheme
%presented here composes a benchmark to quantify the performance
%degradation of these algorithms.
To finalize this section, we note that, while we count on the
secrecy encoding rates to lie in the region specified in
Theorem~\ref{lemma1}, we do not claim any rate in that region is
satisfied by our scheme. In fact, we do not specify the set of
achievable rates for our problem. However, we show that, our scheme
achieves the maximum achievable region of end-to-end rates. The main
significance of the algorithm presented in this section is that, it
can be considered as a benchmark against which all other algorithms
developed with one or more of the assumptions A1.-A3. relaxed can be
compared.

\section{Confidential Multihop Network Control with a Finite Decoding Delay Constraint}
\label{sec:w_delay_const}

%the data is backlogged and is injected in the network at a certain
%rate. We need to clarify that the data is mapped into a block (of
%certain size) of secrecy encoded data, and the receiver can only
%decode it once it receives all the blocks associated with the
%encoded data. Thus, if $N_s=\infty$, the delay is $\infty. For
%finite N_s, delay is finite.

In Section \ref{sec:wo_delay_const}, we propose a dynamic control
algorithm associated with end-to-end secrecy encoding, where
messages are encoded over infinitely many blocks. Hence, the
decoding delay of confidential message may be infinitely long. In
this section, we consider a more practical case by removing
Assumption A1., i.e., there is a hard constraint on the number of
blocks a given confidential message is encoded, $N_s<\infty$. The
entire data including actual confidential bits and the randomization
bits sent by source $s$ is $N_sR_s$, where $R_s$ is defined as
before.
%Note
%that the receiver can decode the message only when it receives all the blocks.
%Thus if $N_s \rightarrow \infty$, we get infinite delay. If $N_s$ is
%finite, delay is finite as well.
We assume that the length of the message $N_sR_s$ is determined a
priori based on the required end-to-end delay between the source and
destination nodes. We also remove Assumption A3., where end-to-end
confidential data rate may not be in general equal to $R_s^{conf}$.
%
%
%Note that even though by encoding over infinitely many blocks one
%may achieve higher long-term private and achperfect secrecy,
%decoding delay may be a more important concern in many practical
%scenarios.
%

Unlike the infinite-block case, since a message is encoded across a
finite number of blocks, subsequent packets associated with a given
secrecy encoded message cannot be decoupled. Therefore, achieving
perfect secrecy for all messages is not possible. Hence, we define
the notion of \textit{secrecy outage}. We say that a secrecy outage
event occurs, when the confidential message is intercepted by any
intermediate node, i.e., the perfect secrecy constraint
\eqref{eq:perfect_privacy} is violated. Secrecy outages can be
completely avoided in the infinite-block scenario, since the network
mechanisms can react to an undesirably large rate of accumulation at
a given node at a time scale faster than the number of blocks across
which the message is encoded. However, here, the reaction time may
be too slow and the accumulated information at a node may already
exceed the threshold for perfect secrecy. Consequently, rather than
perfect secrecy, we impose a constraint on the event that a message
experiences a {\em secrecy outage}. In particular, we assume that,
each source has the knowledge of the amount of accumulated
information at the intermediate nodes for its messages, so it can
identify (but not necessarily avoid) the occurrence of the event of
secrecy outage.%\footnote{In the next section, we address this
%assumption and investigate the possibility of infrequent update of
%the amount of accummulated information in the intermediate nodes.}
On the other hand, unlike the infinite-block case, each of the
messages encoded across finite number of blocks, $k$, can be encoded
with a different confidential rate $R_s^{k,conf}$, determined based
on the history of prior messages experiencing secrecy outages. Thus,
a scheme can adaptively vary its confidential data rate to improve
the performance.

%\footnote{The secrecy outage occurs when the rate of information
%obtained by any intermediate node exceeds the randomization rate,
%i.e., $R_s^k-R_s^{k,conf}$.}

As in Section \ref{sec:wo_delay_const}, our objective is to maximize
the aggregate long-term confidential utility of $K$
source-destination pairs. Let $x_s^p$ be the average rate of
confidential messages injected into the queue of the source node
$s$, $p^{out}_s(R_s^{k,conf})$ be the long-term average of secrecy
outages of the message $k$ of source node $s$ when encoded with
confidentiality rate $R_s^{k,conf}$, and $\gamma_s$ be the maximum
allowable portion of actual confidential bits experiencing secrecy
outage.  We consider the solution of the following optimization
problem:
%\vspace{-0.15in}
\begin{align}
\label{op:objective_func_finite}
&\max_{R_s^{k,conf},  {\cal I}_v(t), {\cal I}_{ji}^s(t)} \sum_{s \in S} U(x_s^p) \\
\label{op:flow_control_finite} \text{s. t.  }   & x_s^p
\leq \bar{R}_s^{conf} \\
\label{op:privacy_outage_finite} & \bar{R}_s^{out} \leq \gamma_s \bar{R}_s^{conf} \\
\label{op:flow_conservation_finite} & \sum_{\{j|(i,j)\in L\}} \bar{\mu}_{ij}^s -  \sum_{\{i|(i,j)\in L\}} \bar{\mu}_{ij}^s \geq 0, \, %\forall \ i \notin S,D \\
\end{align}
%\begin{align}
%\label{op:objective_func_finite} \max &\lim_{T \rightarrow \infty}
%\sum_{s \in S}
%\frac{1}{T}\sum_{t=1}^{T} U(A_s^p(t)) \\
%\label{op:flow_control_finite} \text{s. t.}  & \lim_{T \rightarrow
%\infty}\frac{1}{T} A_s^p(t)
%\leq \lim_{T \rightarrow \infty} \frac{1}{K_s(T)}\sum_{k=1}^{K_s(T)}R_s^p(k) \\
%\label{op:privacy_outage_finite} &\lim_{T \rightarrow \infty}
%\frac{1}{K_s(T)}\sum_{k=1}^{K_s(T)} I\left(
%\sum_{t = t_k^s}^{t_k^d} f_j^{s,k}(t) > B_s-B_{s,T}^p \right) \leq \gamma_s \\
%\label{op:flow_conservation_finite} &\lim_{T \rightarrow \infty}
%\frac{1}{T}\sum_{t=1}^{T} \sum_{e(t)\in E}\sum_{j:(i,j)\in
%e(t)}\mu_{ij}^s(t)  \leq \lim_{T \rightarrow \infty}
%\frac{1}{T}\sum_{t=1}^{T} \sum_{e(t)\in E}\sum_{j:(j,i)\in
%e(t)}\mu_{ji}^s(t)
%\end{align}
%where $K_s(T)$ denotes the number of private packets transmitted by
%the source $s$ in $T$ slots and $f_j^{s,k}(t)$ denotes the
%information leakage to node $j$ about $k^{\text{th}}$ private packet
%of source $s$ in block $t$. Note that $I(a
%> b)$ is indicator variable takes value of one when $a > b$, zero
%otherwise. If $\left( \sum_{t = t_s^{k}}^{t_s^d} f_j^{s,k}(t)
%> B_s-B_{s,T}^p \right)$ for the $k^{\text{th}}$ private packet, we say that the
%intermediate nodes intercept the private message.
where \small $\bar{R}_s^{conf}= \lim_{K \rightarrow
\infty}\frac{1}{K} \sum_{k=1}^K R_s^{k,conf}$ \normalsize and \small
$\bar{R}_s^{out}= \lim_{K \rightarrow \infty}\frac{1}{K}
\sum_{k=1}^K R_s^{k,conf}p^{out}_s(R_s^{k,conf})$. \normalsize

Constraint~\eqref{op:flow_control_finite} ensures that the long-term
service rate is larger than the long-term arrival rate;
Constraint~\eqref{op:privacy_outage_finite} ensures that the portion
of the actual confidential bits experiencing secrecy outage is lower
than $\gamma_s$; and Constraint~\eqref{op:flow_conservation_finite}
is for the flow conservation at intermediate nodes.

\begin{figure}
%\centering
\begin{center}
%\end{tabular}
%\begin{tabular}{c}
 \includegraphics[width=2.5in]{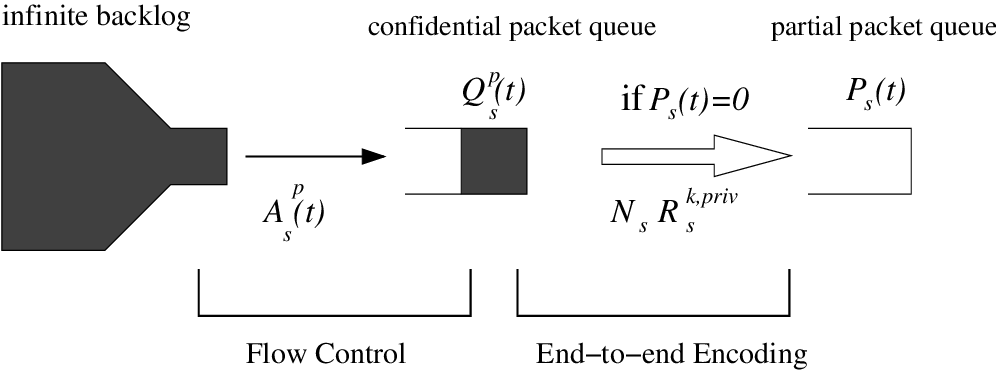}
%(b) Comparison with the passive methods
%\end{tabular}
\vspace{-0.1in} \caption{Queues in a source node used for Control
Algorithm 2.} \vspace{-0.2in}
\label{fig:finite_encoding}  \vspace{-0.1in}     % Give a unique label
\end{center}
\end{figure}

Once again, we employ a cross-layer dynamic control algorithm based
on the stochastic network optimization framework to solve the
optimization problem
(\ref{op:objective_func_finite}-\ref{op:flow_conservation_finite}).
Clearly, in this problem, since the message is encoded across a
finite number of blocks, a control decision given in a time slot
depends on the decisions of the same message given in the subsequent
time slots. Thus, the optimality of dynamic control algorithms
cannot be claimed. In the following, we develop a sub-optimal
solution which performs scheduling by treating the messages as if
they are infinite length messages, but at the same time chooses
confidential encoding rates of individual messages by keeping
account of information experiencing secrecy outages.

The source node $s$ has two separate queues operating at two
different time scales as illustrated in
Fig.~\ref{fig:finite_encoding}. The first queue stores the
confidential information that has been neither secrecy-encoded nor
transmitted in previous blocks. Let $Q_s^p(t)$ denote the length of
the confidential information queue at block $t$. In every block $t$,
$A_s^p(t)$ confidential bits are admitted into the queue, where
$x_s^p$ is the long term average rate of admitted confidential bits.
Departures from this queue occur only when a new secrecy-encoded
message is created. Let $k_s(t)$ be the number of secrecy-encoded
messages created by block $t$.  The $k_s(t)$th confidential message
is encoded with rate $R_s^{k_s(t),conf}$, so the actual confidential
bits in the message is $N_sR_s^{k_s(t),conf}$ whereas the complete
length of the encoded message including the randomization bits is
always $N_sR_s$ for every secrecy-encoded message.  Once a new
confidential message is created, it is admitted to the second queue,
which stores the bits of partially transmitted confidential message
$k_s(t)$. Let $P_s(t)$ denote the size of this partial message queue
at block $t$. The departures from this queue may occur at any block
$t$ depending on the outcome of the scheduling and routing
decisions.  A new secrecy-encoded message is admitted to the queue
only when the partial message queue has emptied, i.e., $P_s(t)=0$.
Hence, $k_s(t+1)=k_s(t)+1$, if $P_s(t)=0$. The evolution of the
states of the queues at each source $s$ can be stated as:

\vspace{-0.1in} \small
\begin{align}
%\label{eq:priv_queue}
\nonumber Q_s^p(t+1) &= \begin{cases} \left[ Q_s^p(t) - N_sR_s^{k_s(t+1),conf} \right]^+ + A_s^p(t) \ \ & \text{if } P_s(t)=0, \\
Q_s^p(t) + A_s^p(t), & \text{otherwise} \end{cases},\\
%\label{eq:partial_queue}
\nonumber P_s(t+1) &= \begin{cases} N_sR_s   \ \ & \text{if } P_s(t)=0  \\
 \left[ P_s(t) - \sum_{i|(s,i)\in L} \mu_{si}^s(t) \right]^+ & \text{otherwise} \end{cases}.
\end{align}
\normalsize
%Let $t_s^k$
%be the block when $k$th new private message is generated by source
%$s$, and $\mathbf{t_s}=\{t_s^k\}$ be the set of blocks when a new
%private message is generated.  Unlike the delay unlimited case,
%source node $s$ encodes each private message possibly with a
%different private encoding rate, $R_s^{k,priv}$.  Also note that a
%new private message is generated only when transmission of
%previously encoded message of length $N_sR_s$ has been completed.
%The second queue stores the bits of an encoded private message which
%is only partially transmitted by block $t$.    However, an arrival to this
%queue occurs only when a new private message is generated, i.e.,
%$k$th arrival occurs at $t_s^k$.
%Let ${\cal D}_s^k(t)$ be indicator
%variable taking value of 1, if $(k-1)$th message is fully
%transmitted by source $s$, i.e., $P_s(t) = 0$.

\noindent At every intermediate node, there is a queue for each
source $s$. Let $Q_i^s(t)$ denote the size of the queue at
intermediate node $j$ for source $s$. Then, we have \vspace{-0.05in}
\begin{align}
Q_i^s(t+1) &= \left[ Q_i^s(t) - \sum_{j|(i,j)\in L}\mu_{ij}^s(t)
\right]^+  + \sum_{j|(j,i)\in L} \mu_{ji}^s(t).\nonumber
\end{align}
%
%\noindent As given \eqref{eq:priv_queue} and
%\eqref{eq:partial_queue}, a departure from private queue and an
%arrival to partial queue occurs only when a new private message is
%generated and previous one is fully transmitted by the source.

In order to determine the occurrence of secrecy outages, each source
keeps track of the accumulated information at each intermediate node
(we address this issue in the next section). Let $Z_i^s(t)$ be the
number of bits that must be accumulated by intermediate node $i$ to
decode the $k_s(t)$th confidential message of source $s$ at block
$t$. Note that a secrecy outage occurs in $k_s(t)$th message, if
$Z_i^s(t)=0$ for any intermediate node $i$.
\vspace{-0.05in} %\footnotesize
\begin{align}
Z_i^{s}(t+1) = \begin{cases} (R_s-R_s^{k_s(t+1),conf}) N_s & \mbox{if } P_s(t)=0\\
\left[ Z_i^{s}(t) - \sum_{j\neq i}f_i^{s,j}(t) \right]^+&
\mbox{otherwise}
\end{cases}.\nonumber
%Z_i^{s}(t+1) = \left[ Z_i^{s}(t) + f_i^{s}(t) -
%\frac{R_s-R_s^{k,conf}}{R_s} \sum_{j|(j,d)\in L, d \in
%D}\mu_{jd}^s(t)\right]^+, \text{if } t \in \mathbf{t_s}
%\label{eq:virtual-info}
\end{align}
%
%\normalsize If $N_s \rightarrow \infty$, the stability of the queue
%given in \eqref{eq:virtual-info} ensures the perfect privacy. For
%finite-block encoding case, the queue is utilized for controlling
%the distribution of the packets transmitted over the paths.
%
The constraint in \eqref{op:privacy_outage_finite} can be
represented by a virtual queue, which keeps account of confidential
information experiencing secrecy outages. Hence, there is only an
arrival of $R_s^{k,conf}$ if $k$th message has undergone secrecy
outage.

\vspace{-0.1in} \small
\begin{align*}
%\label{eq:virtual-privacy}
Y_s^{k+1} = \begin{cases} \left[ Y_s^k+R_s^{k,conf} - \gamma_sR_s^{k,conf} \right]^+ \ \ \text{if outage in $k$th message }\\
\left[Y_s^k - \gamma_s R_s^{k,conf}\right]^+ \ \ \text{if no outage
in $k$th message}
\end{cases}.
\end{align*}

\normalsize Arrivals and departures to this virtual queue are the
number of the confidential bits undergoing secrecy outage and the
number of confidential bits allowed to undergo outage as given by
the outage constraint, respectively. The state of the virtual queue
at any given point is an indicator of the amount by which we have
exceeded the allowable secrecy outage constraint. Thus, the larger
the state of this queue, the more conservative our dynamic algorithm
has to get toward meeting these constraints, i.e., it encodes the
message with less amount of confidential information,
$R_s^{k_s(t),\text{conf}}$ in $k_s(t)$th message.

\normalsize

\noindent {\bf Control Algorithm 2 (with Finite Encoding Block):}\\
\noindent For each source $s$:
\begin{enumerate}
\item[\bf (1)]{\bf End-to-end Encoding:} At every generation of new confidential message, i.e., $P_s(t) = 0$, let $k_s(t+1)=k_s(t)+1$,
and determine end-to-end confidential encoding rate:
\begin{align*}
    R_s^{k_s(t+1),conf} = \argmax_r \left\{ Q_s^p(t)r -Y_s^{k_s(t)} \left(r p_s^{out}(r)-r \gamma_s \right)   \right\}
\end{align*}

\item[\bf (2)] {\bf Flow control:}  At each block $t$, for some $H>0$, each
source $s$ injects $A_s^p(t)$ confidential bits into its queues
\small
\begin{align*}
A_s^p(t)=  \argmax_a \left\{HU_s(a)- Q_s^p(t)a \right\}.
\end{align*}

\normalsize
\item[\bf (3)] {\bf Scheduling:} At each block, $t$ , the scheduler
chooses the set of links $e$ if ${\cal I}_v(t) = 1$ and flow $s$ on
the link $(i,j) \in v$ if ${\cal I }_{ij}^s(t) = 1 $, where

\vspace{-0.15in} \small
\begin{align*}
%&\left({\cal I}_e(t), {\cal I}_{ij}^s(t)\right) = 1\ \forall (i,j)\in e\ \text{and $s$, where} \\
&(s,e)=\argmax_{s\in S,v\in \pazocal{V}}   \left\{ \sum_{(s,i)\in v} \left( \frac{R_s}{R_s^{k_s(t),conf}} Q_s^p(t)+ P_s(t) -Q_i^s(t) \right) \mu_{si}^s(t) \right. \\
&+\left. \sum_{(i,j)\in
v}\left(Q_i^s(t)-Q_j^s(t)\right)\mu_{ij}^s(t) - \sum_{j \in E}
\sum_{i \neq j} Z_j^{s}(t)f_j^{s,i}(t) \right\}
\end{align*}
\normalsize and $Q_s^p(t)$ is multiplied by $R_s/R_s^{k_s(t),conf}$
in order to normalize it to the size of other queues in the network.
\end{enumerate}
Note that the long-term average of secrecy outages $p^{out}_s(R)$
can only be calculated if the scheduling decisions are known a
priori. Since this is not the case,  we use an estimate of secrecy
outage probability as discussed in Section \ref{sec:num_results}.
Note that, $p^{out}_s(R)$ is an increasing function of $R$, since as
$R$ increases, each confidential message is encoded with less
randomization rate, and an intermediate node may intercept the
confidential message with higher probability. Thus, the end-to-end
encoding rate, $R_s^{k_s(t),conf}$ increases as the length of
confidential information queue, $Q_s^p(t)$, increases but it
decreases as the length of the virtual queue, $Y_s^{k_s(t)}$,
increases.

Finally, we would like to re-iterate that, Control Algorithm 2 does
not guarantee obtaining the optimal solution of
\eqref{op:objective_func_finite} -
\eqref{op:flow_conservation_finite} due to the dependance of
decisions between subsequent blocks. We verify by extensive
numerical analysis that its performance is still close to the
optimal in a variety of scenarios.
%\subsection*{Discussion:}
%
%The decision variables of the optimization
%problem become the rate of admitted information from source, $s$, i.e., $x_s$, and the rate allocated to each flow $s$ over
%the link $(i,j)$, i.e., $\mu_{ij}^s$. In addition, the Constraints
%in \eqref{flow_conservation_const_fixed} and
%\eqref{secrecy_const_fixed} are modified as $\sum_{s} \mu_{ij}^s
%\leq R_{ij}$ and $\sum_{\{j|(i,j)\in L\}} \mu_{ij}^s \leq
%(1-\alpha_s) x_s$, respectively.
%%\small
%%\begin{align}
%%\label{objective_func_fixed}
%%    &\max_{x_s, \mu_{ij}^s} \sum_{s\in S} U_s(\alpha_s x_s) \\
%%    \label{stability_const_fixed}
%%    &\text{s.t.  } x_s \leq \sum_{\{i|(s,i)\in L\}} \mu_{si},
%%    \,    \forall s \in S\\
%%    \label{flow_conservation_const_fixed}
%%    & \sum_{s} \mu_{ij}^s \geq R_{ij}, \, \forall \ i \notin S,D \\
%%    \label{secrecy_const_fixed}
%%    & \sum_{\{j|(i,j)\in L\}} \mu_{ij}^s \leq (1-\alpha_s) x_s, \, \forall s \in S, \forall j \notin
%%    S,D ,
%%\end{align}
%%\normalsize
%The resulting optimization problem can be solved by dual
%optimization theory \cite{X_Lin}. By applying similar modifications,
%the algorithm presented in Section \ref{sec:w_delay_const} can be
%applied for wired networks as well. This time, sources adjust their
%end-to-end encoding rates, $R_s^{k,priv}$, according to congestion
%feedbacks as well as their privacy outage histories.

\section{Reducing the Overhead and Distributed Implementation}
\label{sec:implementation}

The algorithms presented in the Section \ref{sec:wo_delay_const} and
\ref{sec:w_delay_const} solve constrained optimization problems in
a centralized fashion. The centralized algorithms provide an upper
bound on the network performance, which can be used a benchmark to
evaluate the performance of distributed algorithms. In this section,
we design algorithms relaxing the assumptions necessary for Control
Algorithm 1, where the instantaneous queue length information is not
available and/or a centralized scheduler is absent in the network.
Note that the flow control portions of the algorithms provided in the previous sections were already distributed, i.e, each node decides its admitted flow by only local information. Thus, they remain the same as given in Section~\ref{sec:wo_delay_const}, in the rest of the sequel.

\subsection{Infrequent Queue Length Updates}
\label{sec:infrequent}

%In this subsection,  we show that it is easy to incorporate
%infrequent queue length information that can simplify the
%implementation of our algorithm. In order to better understand the
%amount of queue length information (containing real and virtual
%queues) required to operate a scheduling algorithm, consider a
%wireless multi-hop network where users being scheduled to transmit
%by a centralized entity. To perform scheduling, each user would have
%to quantize and encode its own queue length and transmit it to the
%scheduler, along with the payload. This would require a variable
%number of bits per time slot depending on the longest queue.
%Further, since the queue lengths are unbounded, there is a need for
%a dynamic quantization scheme that necessitates further
%co-ordination between the nodes and the scheduler. Thus,
In this section, we consider the setup described in
Section~\ref{sec:wo_delay_const}, and relax the assumption of the
availability of the queue state information of each node at every
point in time. Indeed, scheduling requires a significant overhead
due to control traffic carrying the queue length information across
the entire network. To reduce this overhead, we consider
transmission of queue length information not in every slot, but once
every $K$ slots. Let $\hat{Q}_s(t)$ and $\hat{Q}_i^s(t)$ denote the
estimates of the queue lengths at source $s$ and at node $i$ for
commodity $s$, respectively, at time $t$. Furthermore, let
$\hat{Z}_i^s(t)$ denote the estimate of the virtual queue at node
$i$ used for the accumulated information about commodity $s$. In
particular, these estimates are the last updates of the queue
lengths, prior to time $t$, received by the scheduler. Further,
suppose that at each time slot, the scheduler gives the routing
decision according to the solution of the following equation:

\vspace{-0.2in}\small
\begin{align*}
%&\left({\cal I}_e(t), {\cal I}_{ij}^s(t)\right) = 1\ \forall (i,j)\in e\ \text{and $s$, where} \\
&(s,v)=\argmax_{s\in S,v\in \pazocal{V}} \left\{ \sum_{(i,j)\in
v}(\hat{Q}_i^s(t)-\hat{Q}_j^s(t))\mu_{ij}^s(t) - \sum_{j \in E}
\sum_{i \neq j} \hat{Z}_j^s(t)f_j^{s,i}(t) \right\} .
\end{align*}
\normalsize

Next, we show that the performance attained by a system where queue
lengths are updated infrequently is again arbitrarily close to the
optimal solution.

\begin{theorem}
\label{thm:optimalcontrol-2}
 If $R_{ij}(t)<\infty$ for all $(i,j)$ links and for all $t$ blocks, then control
algorithm satisfies:

\small \vspace{-0.2in}
\begin{align*}
\liminf_{t\rightarrow\infty}\frac{1}{t}\sum_{\tau=0}^{t-1}\sum_{s\in
S}
\E{U_s(\tau)} &\geqslant U^* - \frac{B+B'(K-1)}{H} \\
\limsup_{t\rightarrow\infty}\frac{1}{t}\sum_{\tau=0}^{t-1}\sum_{s\in
S}
\E{Q_s(\tau)} &\leqslant \frac{B+B'(K-1)+H(\bar{U}-U^*)}{\epsilon_1'} \\
\limsup_{t\rightarrow\infty}\frac{1}{t}\sum_{\tau=0}^{t-1}\sum_{s\in
S}\sum_{i \notin S,D}
\E{Q_i^s(\tau)} &\leqslant \frac{B+B'(K-1)+H(\bar{U}-U^*)}{\epsilon_2'} \\
\end{align*}

\normalsize\vspace{-0.2in} \noindent where
$B,B',\epsilon_1',\epsilon_2'$ are positive constants, $U^*$ is the
optimal aggregate utility, i.e., the solution of
(\ref{objective_func_fixed}-\ref{secrecy_const_fixed}), and
$\bar{U}$ is the maximum possible instantaneous aggregate utility.
\end{theorem}

The proof of Theorem \ref{thm:optimalcontrol-2} is given in Appendix
\ref{proof:optimalcontrol-2}. This theorem shows that it is still
possible to get arbitrarily close to the optimal utility by choosing
$H$ sufficiently large. However, the lack of availability of timely
queue state information negatively affects the performance bounds.
In particular, for a given value of $H$, the bound on the achieved
utility decreases by a factor, proportional to $K$. Likewise, the
upper bounds on the queue sizes increase by a factor, proportional
to $K$. Although similar results were obtained in \cite{A_Eryilmaz}
and \cite{Manikandan} as well, here we show that confidentiality is
still achieved with the the infrequent periodic update of the queue
lengths.
%In other words, while the infrequent periodic update of the queue lengths is close to optimal depending on $H$ value, it may result in poor delay performance.
%Another alternative to update queue lengths other than periodic
%update is to update the queue length information of each queue
%whenever the absolute value of the difference between the current
%length and the last update exceeds some threshold. Along the lines
%of the proof of Theorem \ref{thm:optimalcontrol-2}, this algorithm
%can be shown to be arbitrarily close to optimal. In addition, it was
%shown in~\cite{A_Eryilmaz} that this update mechanism reduces the
%average queue sizes as compared to the periodic sampling.

%Finally, we note that delayed queue length updates can also be cast
%in the same framework as above

\subsection{Distributed Implementation}
\label{sec:distributed}

In the previous section, we dealt with the issue of reducing the
overhead of the centralized scheduling. Here, we seek a distributed
algorithm where each node participates in scheduling using only
local information: We assume that the nodes have information of the
instantaneous CSI only between themselves and their neighbors, and
only of the queue lengths of their neighbors.

\setcounter{algorithm}{2}
\begin{algorithm}
\caption{Distributed Scheduling Algorithm} \label{dist_algorithm}
%\begin{algorithmic}
Each node $i$ carries out the following steps over each block $t$:

1) Calculate weight $W_{ij}^s(t)
=(Q_i^s(t)-Q_j^s(t))R_{ij}(t)-\sum_{k \in E} Z_k^s(t)R_{ik}(t)$ for
each link pair $(i,j)$. Ties are broken randomly.

2) Find node $j^*$ such that $W_{ij^*}^s(t)$ is maximized over all
links $(i,j)$ with free neighbors $j$. If having received a matching
request from $j^*$, then link $(i,j)$ is a matched link. Node $i$
sends a matched reply to $j^*$ and a drop message to all other free
neighbors. Otherwise, node $i$ sends a matching request to node
$j^*$.

3) Upon receiving a matching request from neighbor $j$, if $j =
j^*$, then link $(i,j)$ is a matched link. Node $i$ sends a matched
reply to node $j$ and a drop message to all other free neighbors. If
$j \neq j^*$, node $i$ just stores the received message.

4) Upon receiving a matched reply from neighbor $j$, node $i$ knows
link $(i,j)$ is a matched link, and sends a drop message to all
other free neighbors.

5) Upon receiving a drop message from neighbor $j$, node $i$ knows
that $j$ is in a matched link, and excludes $j$ from its set of free
neighbors.

6) If node $i$ is in a matched link or has no free neighbors, no
further action is taken. Otherwise, it repeats steps 2) through 5).

7) Matched links are allowed to transmit, i.e., if link $(i,j)$ is a
matched link, node $i$ transmits data to node $j$.
%\end{algorithmic}
\end{algorithm}

The scheduling problem of the control algorithms designed in the
previous sections can be reduced to a maximum weighted matching
problem, which is polynomial time solvable, but requires a
centralized implementation.\footnote{A matching in a graph is a
subset of links, no two of which share a common node. The weight of
a matching is the total weight of all its links. A maximum weighted
matching in a graph is a matching whose weight is maximized over all
matchings of the graph.} Each node needs to notify the central node
of its weight and local connectivity information such that the
central node can reconstruct the network topology. A few distributed
approximation algorithms exist for the maximum weighted matching
problem, e.g., \cite{Sanghavi} and \cite{Hoepman}. Here, we make use
of the distributed scheduling algorithm presented in \cite{Hoepman},
where the maximum weighted matching is obtained sequentially. Let a
link that has been chosen to be in the matching be called a matched
link. Nodes that are not related to any matched link are called
free. A matching request is transmitted to enquire the possibility
to choose the link with a neighbor as a matched link. A matched
reply is sent to confirm that the link with a neighbor is matched. A
node sends a drop message to inform its neighbors that it is not
free anymore. Algorithm \ref{dist_algorithm} gives the details of
the distributed scheduling algorithm.

%the above algorithm has linear complexity, i.e., $O(L)$, which is
%important for sacability and complexity

%Note that  the algorithm is locally optimal. The reason is that
%define a link $(i,j)$ to have locally the maximum weight, its weight
%is maximized over all links with free neighbors. We can see that
%this algorithm selects locally heaviest links as matched. However,
%local optimality do not converse to global optimality.

At the beginning of each slot, node $i$ does not have information
about which of its neighbors will transmit in that slot. Thus, it
considers as leaked information to all its free neighbors while
computing weights of its links, even if some of those neighbors may
transmit in the following iteration.\footnote{As we consider a
node-exclusive interference model, nodes cannot transmit and receive
information at the same time.} However, when all matched links are
set, node $i$ overhears the surrounding transmissions, and updates
the virtual queues related to the overheard information of its
neighbors, i.e., $Z_j^s(t)$. In Section \ref{sec:num_results}, we
demonstrate that the proposed distributed scheduling algorithm
results in a small degradation in the overall performance.

\section{Numerical Results}
\label{sec:num_results}

%\vspace{-0.3in}

\begin{figure*}[htp]
\centerline{
%\subfloat[Arrival and service rates]{\includegraphics[width=1.9in]{./Figures_SFC/rates_vs_V.eps}%
%\label{fig:num1}} \hfil
\subfloat[$\alpha_1$ vs Long-Term Confidential Data Rate $(\alpha_sx_s)$]{\includegraphics[width=2.3in]{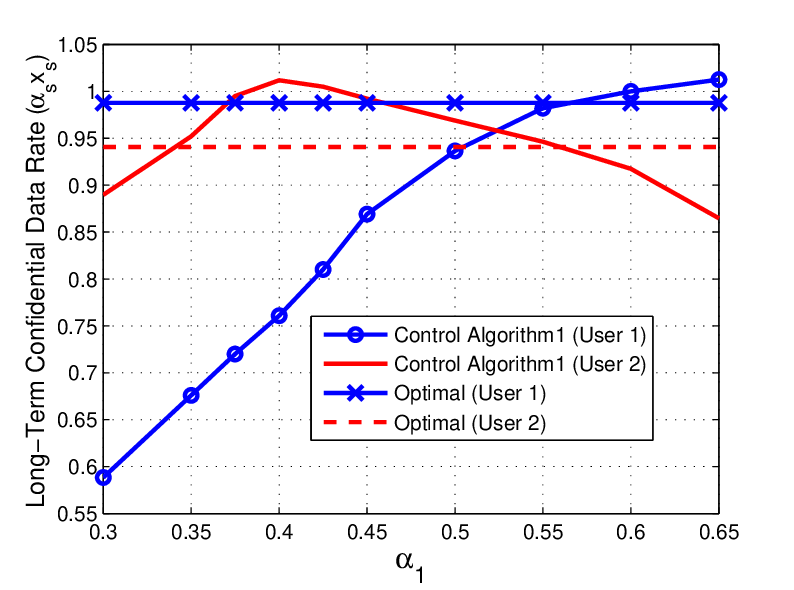}%
\label{fig:alpha_rate}} \hfil
\subfloat[$\alpha_1$ vs Long-term Total Utility]{\includegraphics[width=2.3in]{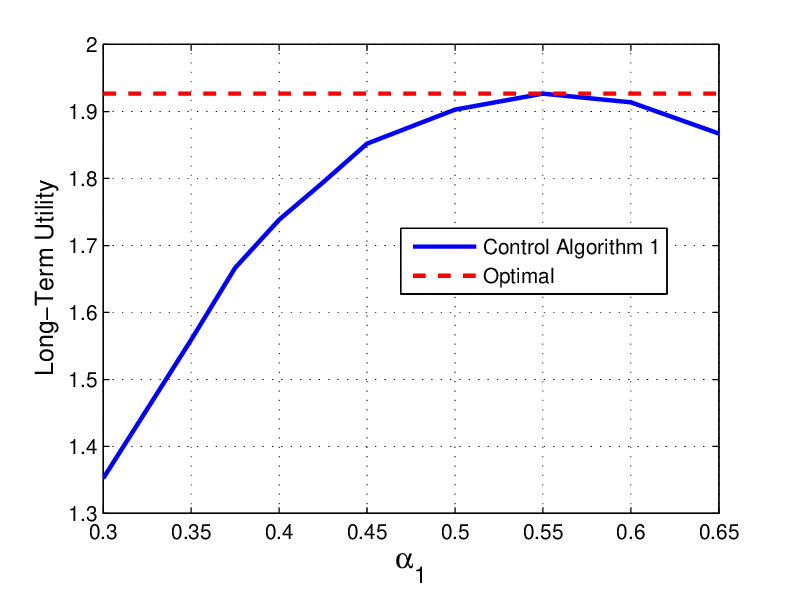}%
\label{fig:alpha_utility}} } \caption{Performance evaluation of
Control Algorithm 1 presented in Section \ref{sec:wo_delay_const},
when all intermediate nodes are eavesdroppers. } \vspace{-0.2in}
\end{figure*}

\begin{figure*}[htp]
\centerline{
%\subfloat[Arrival and service rates]{\includegraphics[width=1.9in]{./Figures_SFC/rates_vs_V.eps}%
%\label{fig:num1}} \hfil
\subfloat[$\alpha_1$ vs Long-Term Confidential Data Rate $(\alpha_sx_s)$]{\includegraphics[width=2.3in]{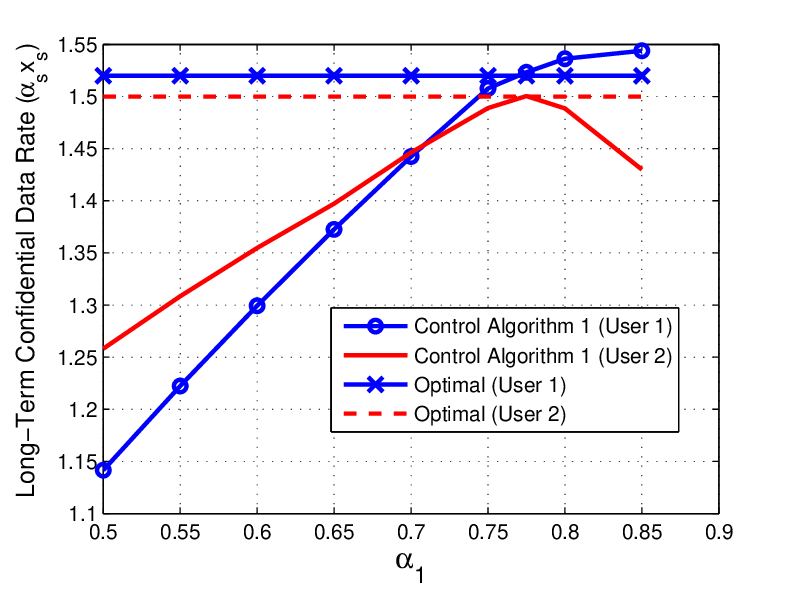}%
\label{fig:alpha_rate2}} \hfil
\subfloat[$\alpha_1$ vs Long-term Total Utility]{\includegraphics[width=2.3in]{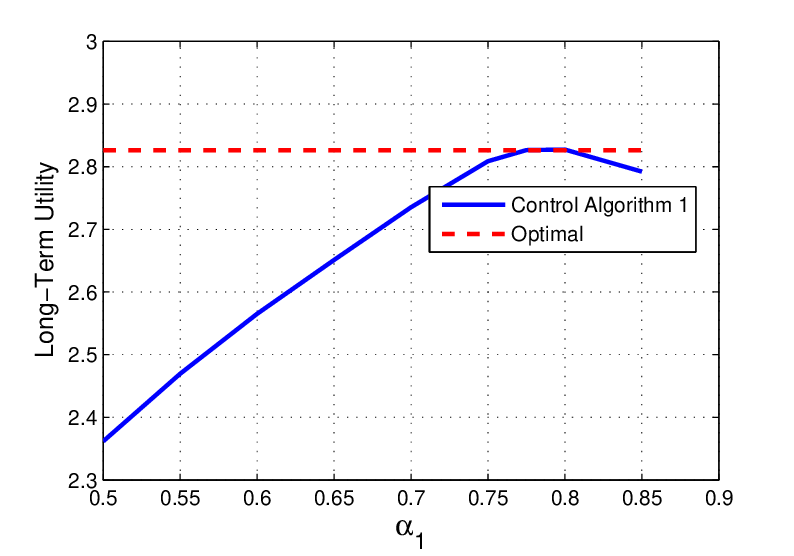}%
\label{fig:alpha_utility2}} } \caption{Performance evaluation of
Control Algorithm 1 presented in Section \ref{sec:wo_delay_const},
when the number of eavesdroppers among all intermediate nodes are
two. } \vspace{-0.2in}
\end{figure*}

%In our numerical experiments, we have considered the network
%depicted in Fig. \ref{fig:independent_set}.
The channels between nodes are modeled as iid Rayleigh fading
Gaussian channels. The noise normalized transmit power is taken as
constant and identical to $P=1$ in every block and for all nodes.
Let the power gain of the channel between nodes $i$ and $j$ be
$h_{ij}(t)$ at block $t$. Then, as $N_1 \rightarrow \infty$,
$R_{ij}(t) = \log(1+Ph_{ij}(t))$. The power gains of the channels
are exponentially distributed, where the means of the link are as
given in Table I. We consider a logarithmic utility function,
$U_s(t) = \kappa + \log(A_s^p(t))$, where $A_s^p(t)$ is the
confidential information admitted in block $t$.\footnote{We utilize
logarithmic utility function to provide proportional fairness.} Note
that, $A_s^p(t) = \alpha_s A_s(t)$ for the control algorithm
presented in Section \ref{sec:wo_delay_const}. We take $\kappa = 3$
and $H = 100$ in all experiments.

%\vspace{-0.05in}

%\begin{table}[h]
%\label{table:mean_channel} \caption{Mean channel gains} \scriptsize
%\vspace{-0.2in}
%\begin{center}
%    \begin{tabular}{ | c | c | c | c | c | c | c | c |}
%    \hline
%    ($s_1$, 1) & ($s_1$, 2) & ($s_2$, 2) & ($s_2$, 3) & (1, $d_1$)& (2, $d_1$)& (2, $d_2$)& (3, $d_2$) \\ \hline
%    6 & 8 & 10 & 4 & 8 & 6 & 4 & 6  \\ \hline
%    \end{tabular}
%\end{center}
%\normalsize \vspace{-0.2in}
%\end{table}

\begin{table}[h]
\label{table:mean_channel1} \caption{Mean channel gains} \scriptsize
\vspace{-0.2in}
\begin{center}
    \begin{tabular}{ | c | c | c | c | c | c | c | c |}
    \hline
    ($s_1$, 1) & ($s_1$, 2) & ($s_1$, 3) & ($s_2$, 2) &  ($s_2$, 3) & ($s_2$, 4)  \\ \hline
    6 & 8 & 10 & 4 & 8 & 6  \\ \hline
    (1, $d_1$)& (2, $d_1$)& (3, $d_1$)& (2, $d_2$)& (3, $d_2$) & (4,
    $d_2$) \\ \hline
    6 & 8 & 10 & 4 & 8 & 6 \\ \hline
    \end{tabular}
\end{center}
\normalsize \vspace{-0.2in}
\end{table}

In Fig. \ref{fig:alpha_rate}-\ref{fig:alpha_utility}, we investigate
the performance of Control Algorithm 1, when all intermediate nodes
are considered as eavesdroppers, i.e., $\{1,2,3,4\} \in E$. For the
network depicted in Fig. \ref{fig:independent_set}, we numerically
obtain the encoding rate, $\alpha_s^*$, resulting in the maximum
long-term total utility for source $s$. For the above set of values,
we obtain $\alpha_1^* = 0.55$ and $\alpha_2^* = 0.509$, which
corresponds to optimal long-term average arrival rates $x_1^* = 1.8$
and $x_2^* = 1.85$, respectively. In the experiments, we fix
$\alpha_2 = \alpha_2^*$ and vary the value of $\alpha_1$ to analyze
the effect of $\alpha_s$ on the confidential data rates and total
utility. From Fig.~\ref{fig:alpha_rate}, we first notice that,
long-term confidential data rate of source $s_1$ increases with
increasing $\alpha_1$, since source $s_1$ sends a larger amount of
confidential information for each encoded message. It is interesting
to note that long-term confidential data rate of source $s_2$
increases initially with increasing $\alpha_1$. This is because, for
low $\alpha_1$ values, in order to provide fairness between the
sources, source $s_1$ admits more packets to its queue (e.g., $x_1 =
1.96$, when $\alpha_1 = 0.3$), increasing its queue size. As a
result, scheduling decisions are dictated by the stability
constraint of source $s_1$'s queue, and thus, the long-term arrival
rate of source $s_2$ is lower when $\alpha_1$ is smaller (e.g., $x_2
= 1.78$ when $\alpha_1 = 0.3$). However, when $\alpha_1$ is high,
satisfying the perfect secrecy dominates the scheduling decisions,
and source $s_1$ divides its transmission over the paths more
equally at the expense of lower long-term arrival rates, $x_1$ and
$x_2$. %This is the reason why we observe a decrease in long-term
%privacy rate of source $s_2$ when $\alpha_1$ is high.
Fig. \ref{fig:alpha_utility} depicts the relationship between
$\alpha_1$ and the long-term total utility. As expected, the total
utility increases with increasing $\alpha_1$ until $\alpha_1 =
\alpha_1^*$. As $\alpha_1$ increases, there is a gain due to
incorporating more confidential information into each encoded
message of source $s_1$. However, when $\alpha_1$ is high, long-term
arrival rates of both sources decrease as discussed previously.
Thus, when $\alpha_1
> \alpha_1^*$, the loss due to decrease in $x_1$ and $x_2$
dominates the gain due to increasing $\alpha_1$.

Next, in Fig. \ref{fig:alpha_rate2}-\ref{fig:alpha_utility2}, we
conducted the same analysis, when the number of eavesdroppers among
all intermediate nodes is 2, i.e., the cardinality of set $E$ is 2.
We run simulations for all such possible sets of $E$, i.e.,
$\{1,4\}\in E$ or $\{2,3\}\in E$, and the results in Fig.
\ref{fig:alpha_rate2}-\ref{fig:alpha_utility2} are the average of
the rates obtained for each possible set of $E$. We obtain the
average encoding rates as $\alpha_1^* = 0.775$ and $\alpha_2^* =
0.77$, which corresponds to long-term average arrival rates $x_1^* =
1.96$ and $x_2^* = 1.95$, respectively. Again, we fix $\alpha_2 =
\alpha_2^*$ and vary the value of $\alpha_1$ to analyze the effect
of $\alpha_s$ on the confidential data rates and total utility.
Comparing Fig. \ref{fig:alpha_rate2}-\ref{fig:alpha_utility2} and
Fig. \ref{fig:alpha_rate}-\ref{fig:alpha_utility}, we note that  the
long-term confidential data rate increases as the number of
eavesdroppers among the intermediate nodes decreases. This is
because there are paths without eavesdroppers which can be
facilitated to encapsulate more confidential information in each
encoded message. Lastly, we investigate the same scenario with no
attackers in the network. In this scenario, Control Algorithm 1
simply corresponds to the backpressure algorithm, since we can send
the confidential information without encoding, i.e., with
$\alpha_1^* = 1$ and $\alpha_2^* = 1$, and with the arrival rates
$x_1 = 1.98$ and $ x_1 = 1.92$.

\begin{figure}[t]
%\centering
\begin{center}
%\end{tabular}
%\begin{tabular}{c}
 \includegraphics[width=1.2in]{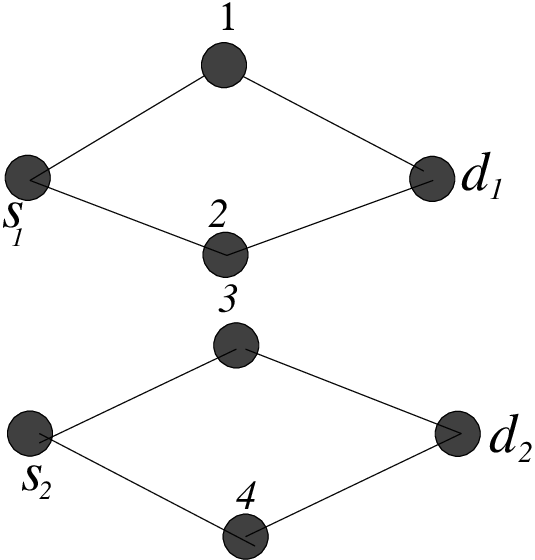}
%(b) Comparison with the passive methods
%\end{tabular}
 \caption{A multi-hop
network with two available paths.}%\vspace{-0.1in}
\label{fig:independent_set2}       % Give a unique label
\end{center}
\vspace{-0.3in}
\end{figure}

\begin{figure*}[htp]
\centerline{
%\subfloat[Arrival and service rates]{\includegraphics[width=1.9in]{./Figures_SFC/rates_vs_V.eps}%
%\label{fig:num1}} \hfil
\subfloat[$N_sR_s$ vs Long-Term Confidential Data Rate $(x_s^p)$]{\includegraphics[width=2.3in]{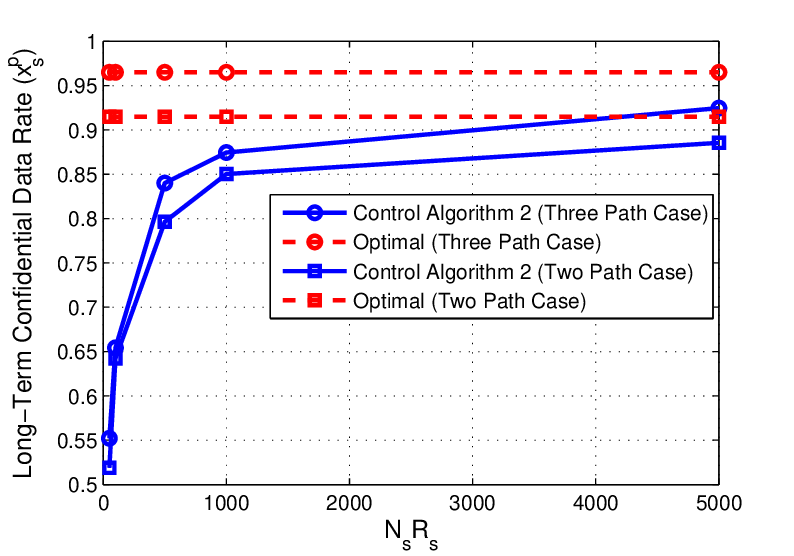}%
\label{fig:rate_Bs}} \hfil
\subfloat[$N_sR_s$ vs Delay (Average number of slots)]{\includegraphics[width=2.3in]{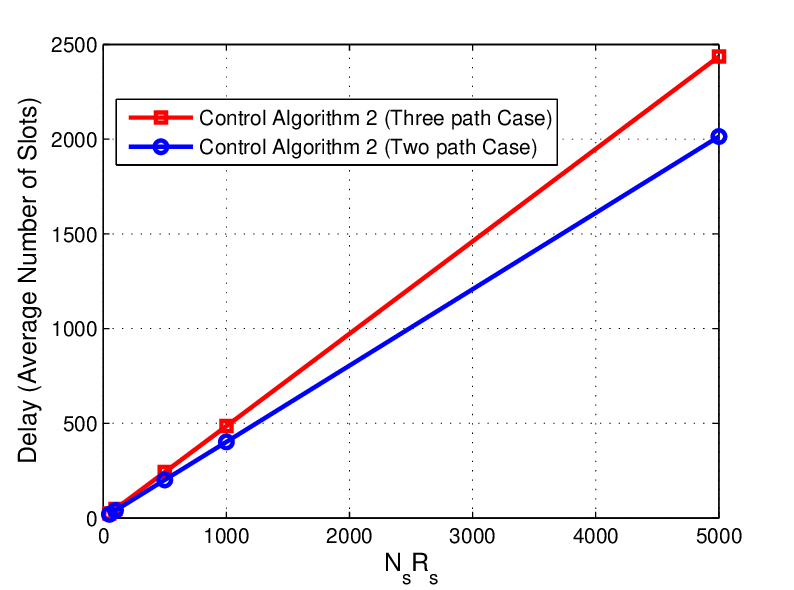}%
\label{fig:delay_Bs}}
\subfloat[$\gamma$ vs Long-Term Confidential Data Rate $(x_s^p)$]{\includegraphics[width=2.3in]{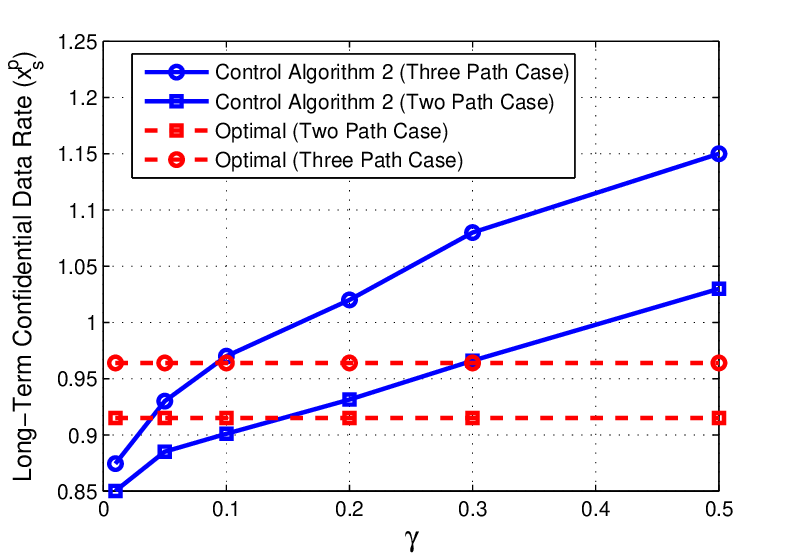}%
\label{fig:gamma_Bs}}}

\caption{Performance evaluation of Control Algorithm 2 presented in
Section \ref{sec:w_delay_const}.} \vspace{-0.2in}
\end{figure*}

We next analyze the performance of Control Algorithm 2 with varying
number of available paths. In particular, we investigated those
networks with two and three paths, as given in Fig.
\ref{fig:independent_set2} and Fig. \ref{fig:independent_set},
respectively. Note that in the network in Fig.
\ref{fig:independent_set2}, nodes $s_1$ and $s_2$ do not have a
communication link to nodes $3$ and $2$. For the network in Fig.
\ref{fig:independent_set2}, we use the main channel gains given in
Table I except $(s_1,3), (s_2,2)$ and $(3,d_1),(2,d_2)$ which have
zero mean channel gains. In numerical experiments, we use
$\frac{R^2}{(R_s)^2}$ as the estimate of $p^{out}_s(R)$ for
$0<R<R_s$. In Fig. \ref{fig:rate_Bs}, the effect of increasing
$N_sR_s$ on the long-term confidential data rate, $x_s^p$, is shown.
We first take the secrecy outage parameter, $\gamma_s = 0.01$, for
all users. Fig. \ref{fig:rate_Bs} depicts that when $N_sR_s = 50$,
the long-term confidential data rate is reduced by approximately
50\%, compared to the optimal rates obtained for $N_s \rightarrow
\infty$.\footnote{If the average rate in a block $t$ is 0.5
bits/channel use, the message is approximately transmitted in 100
blocks.} However, the confidential data rate increases with
increasing $N_sR_s$, and it gets close to the optimal confidential
data rates, $\alpha_s^*x_s^*$, when $N_sR_s$ is large enough, i.e.,
$N_sR_s = 5000$. This is due to fact that when the transmission of a
message takes smaller number of blocks, the portion of confidential
bits inserted into the codeword, $R_s^{k,conf}/R_s$, gets smaller to
satisfy the secrecy constraint. In addition, as the $N_sR_s$
increases, the dependance of the scheduling decisions between
subsequent packets associated with a given secrecy-encoded message
decreases, so i.i.d. approximation of control algorithm presented in
Section \ref{sec:w_delay_const} becomes more accurate. In Fig.
\ref{fig:delay_Bs}, we investigate the end-to-end delay with
increasing $N_sR_s$. The end-to-end delay is the average number of
slots used to transmit an encoded confidential message from the
source to its destination. Note that even though  long-term
confidential data rate increases with an increasing number of
available paths, it results in higher delay as depicted in Fig.
\ref{fig:delay_Bs}. The reason is that with more paths we can encode
the message with more confidential information, but this may cause
congestion among the shared links. Finally, we investigate the
effect of secrecy outage parameter, $\gamma_s = \gamma$ for all
sources, on $x_s^p$. Fig.~\ref{fig:gamma_Bs} shows that when the
secrecy outage constraint is relaxed, i.e., $\gamma$ is increased,
the long-term confidential data rate increases for both networks.
This result is expected, since sources can insert more confidential
information into the encoded message, $R_s^{k,conf}$, with a higher
secrecy outage parameter. Note that, the optimal rate is obtained
when there is no secrecy outages. Thus, for $\gamma \geq 0.1$ and
$\gamma \geq 0.15$, the long-term confidential data rates exceed the
optimal rates for the network in Fig. \ref{fig:independent_set} and
\ref{fig:independent_set2}, respectively.

%\begin{figure*}[htp]
%\centerline{
%\subfloat[Arrival and service rates]{\includegraphics[width=1.9in]{./Figures_SFC/rates_vs_V.eps}%
%\label{fig:num1}} \hfil
%\subfloat[$K$ vs Long-Term Confidential Data Rate $(\alpha_sx_s)$]{\includegraphics[width=2.3in]{ratevsK_journal2.eps}%
%\label{fig:K_rate}} \hfil
%\subfloat[$K$ vs Average Queue Size]{\includegraphics[width=2.3in]{queuesizevsK.eps}%
%\label{fig:K_queue_size}} } \caption{Performance evaluation of
%infrequent queue update algorithm presented in Section
%\ref{sec:infrequent}.}
%\end{figure*}

\begin{figure*}[htp]
\centerline{
%\subfloat[Arrival and service rates]{\includegraphics[width=1.9in]{./Figures_SFC/rates_vs_V.eps}%
%\label{fig:num1}} \hfil
\subfloat[$K$ vs Long-Term Confidential Data Rate $(\alpha_sx_s)$]{\includegraphics[width=2.3in]{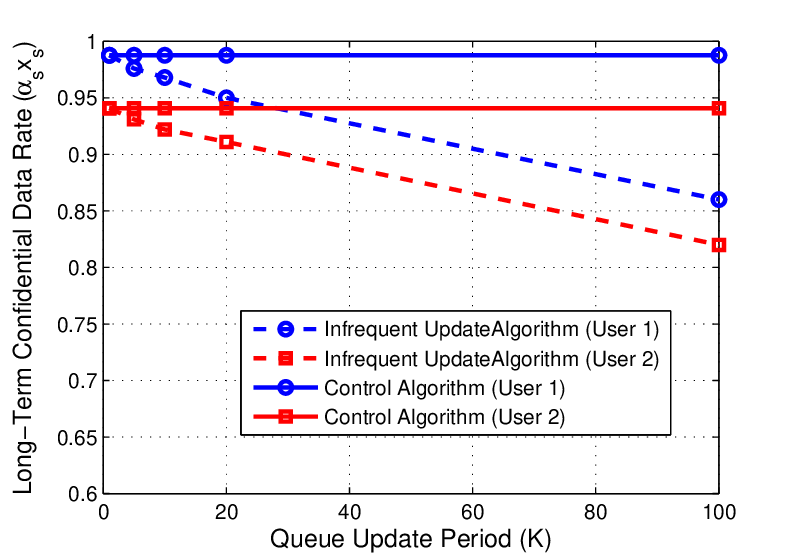}%
\label{fig:K_rate}} \hfil
\subfloat[$\alpha_1$ vs Long-Term Confidential Data Rate $(\alpha_sx_s)$]{\includegraphics[width=2.3in]{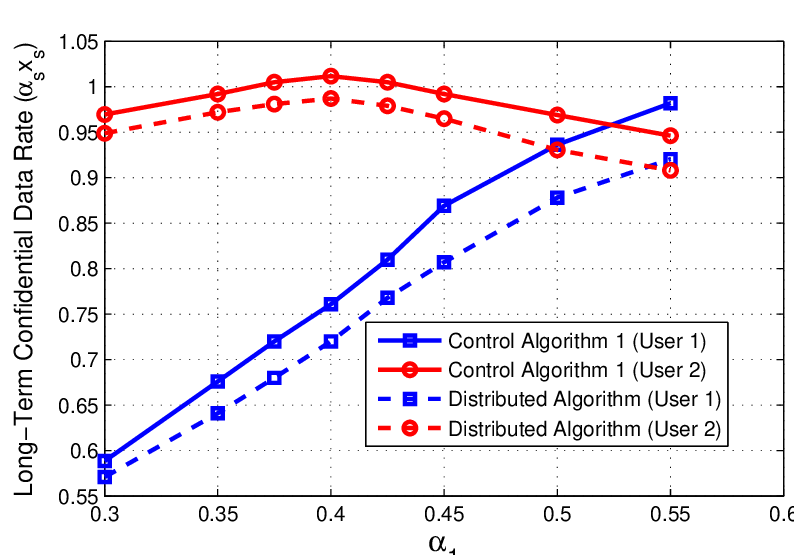}%
\label{fig:alpha_rate_dist}} \hfil
\subfloat[$\alpha_1$ vs Long-term Total Utility]{\includegraphics[width=2.3in]{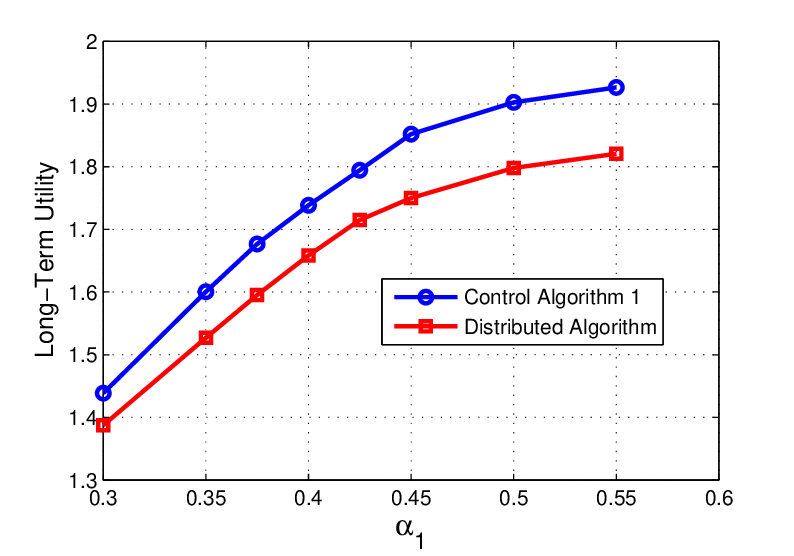}%
\label{fig:alpha_utility_dist}} } \caption{Performance evaluation of
queue update and distributed algorithms presented in Section
\ref{sec:implementation}.}
\end{figure*}

Next, we analyze the performance of the algorithm presented in
Section \ref{sec:infrequent} for the network depicted in Fig.
\ref{fig:independent_set}, where the queue length information is
periodically updated (we refer to this policy as the Infrequent
Update Algorithm). We first analyze the effect of the periodic
update parameter, $K$, on the long-term confidential data rate when
$H$ is the same for all experiments, i.e., $H=100$. In addition,
$\alpha_1$ and $\alpha_2$ are selected as 0.55 and 0.509,
respectively. In Fig. \ref{fig:K_rate}, we observe that as expected,
the confidential data rate decreases with increasing $K$. This is
due to fact that with a large $K$, the algorithm cannot closely
track the queue length values which in turn deteriorates the
performance of the scheduler. We evaluate the effect of $K$ on the
average queue size, when both infrequent update algorithm and
Control Algorithm 1 converge to the optimal point fairly closely. To
achieve near-optimal performance, we set the value of $H$ for
different values of $K$ leading to optimal confidential data rates.
In simulations, we observed that with increasing $K$, we obtain
larger average queue sizes, i.e., %the packets experience
larger delays, confirming theoretical results.
%In Fig.
%\ref{fig:K_queue_size}, we plot average queue sizes with respect to
%$K$. Confirming theoretical results, with increasing $K$, we obtain
%larger average queue sizes, i.e., %the packets experience
%larger delays.

Finally, we analyze the performance of control algorithm presented
in Section \ref{sec:distributed}, where distributed scheduling is
performed (We refer to this policy as Distributed Algorithm). The
results are shown in Fig. \ref{fig:alpha_rate_dist} and Fig.
\ref{fig:alpha_utility_dist}. Here, we select $\alpha_2 = 0.509$ and
varies $\alpha_1$. As expected, the long-term confidential data
rates are smaller than those achieved with centralized scheduling,
since topology and queue length information is not known globally,
and the routing pattern is changed due to distributed scheduling.
Our simulation results show that the degradation of performance of
Control Algorithm 1 with distributed scheduling is relatively small:
distributed scheduling results in a approximately 10\% reduction in
aggregate utility.

%Combined with its low communication overhead, fast convergence, and
%good performance with distributed scheduling, our cross-layer design
%scheme is promising for practical implementation.

%Table II summarizes the routing pattern has been changed, due to the
%distributed scheduling. Also note that every link is used in
%routing, since each link has a chance to be a locally heaviest link.
%Though its worst case performance bound is 1=2,

\vspace{-0.1in}
\section{Conclusion}

%In this paper, we studied multi-hop network control under secrecy
%constraint.

In this paper, we considered the problem of resource allocation in
wireless multi-hop networks where sources have confidential
information to be transmitted to their corresponding destinations
with the help of intermediate nodes over time-varying uplink
channels. All intermediate nodes are considered as internal
eavesdroppers from which the confidential information needs to be
protected. To provide confidentiality in such setting, we propose
encoding the message over long blocks of information which are
transmitted over different paths. Then, we designed a dynamic
control algorithm for a given encoding rate and we prove that our
algorithm achieves utility arbitrarily close to the maximum
achievable utility. In this problem, we find out that increasing the
flow rate and keeping confidentiality is two conflicting objective
unlike standard dynamic algorithms, and the proposed algorithm also
considers spatial distribution of the flows over each path. Next, we
%relax the assumption of encoding messages over infinite blocks, and
consider the system, where the messages are encoded over finite
number of blocks. For this system, transmissions of each block of
the same message are dependant with each other. Thus, we propose a
sub-optimal algorithm, and show that the proposed algorithm
approaches the optimal solution as the number of blocks which the
message are encoded, increases.

Finally, we deal with implementational issues of the proposed
algorithms. First, we decrease overhead imposed by the updates
transmitted to the scheduler. For that purpose, we design infrequent
queue update algorithm, where users updates their queue length
information periodically. We show that this algorithm again
approaches the optimal solution in the expense of increasing average
queue lengths. Then, we investigate distributed version of our
dynamic control algorithms, where the scheduler decision is given
according to local information available to each node. The
simulation results illustrated that the reduction in confidentiality
rate due to usage of distributed algorithm is relatively small.

\bibliographystyle{IEEEtran}
\bibliography{macros,biblio,macros_abbrev,newsecurity}

\begin{thebibliography}{10}
\providecommand{\url}[1]{#1}
\csname url@samestyle\endcsname
\providecommand{\newblock}{\relax}
\providecommand{\bibinfo}[2]{#2}
\providecommand{\BIBentrySTDinterwordspacing}{\spaceskip=0pt\relax}
\providecommand{\BIBentryALTinterwordstretchfactor}{4}
\providecommand{\BIBentryALTinterwordspacing}{\spaceskip=\fontdimen2\font plus
\BIBentryALTinterwordstretchfactor\fontdimen3\font minus
  \fontdimen4\font\relax}
\providecommand{\BIBforeignlanguage}[2]{{%
\expandafter\ifx\csname l@#1\endcsname\relax
\typeout{** WARNING: IEEEtran.bst: No hyphenation pattern has been}%
\typeout{** loaded for the language `#1'. Using the pattern for}%
\typeout{** the default language instead.}%
\else
\language=\csname l@#1\endcsname
\fi
#2}}
\providecommand{\BIBdecl}{\relax}
\BIBdecl

\bibitem{Georgiadis}
L.~Georgiadis, M.~J. Neely, and L.~Tassiulas, ``{Resouce Allocation and
  Cross-layer Control in Wireless Networks},'' \emph{Foundations and Trends in
  Networking}, vol.~1, no.~1, pp. 1--144, 2006.

\bibitem{Xiaojun}
X.~Lin, N.~B. Shroff, and R.~Srikant, ``{On the Connection-Level Stability of
  Congestion-Controlled Communication Networks},'' \emph{IEEE Trans. Inf.
  Theory}, vol.~54, no.~5, pp. 2317--2338, May 2008.

\bibitem{Y_Chen}
Y.~Chen, R.~Hwang, and Y.~Lin, ``Multipath qos routing with bandwidth
  guarantee,'' in \emph{Proc. IEEE Global Telecommunications Conf.}, vol.~4,
  San Antonio, TX, Sep. 2001, pp. 2199--2203.

\bibitem{X_Lin}
X.~Lin and N.~B. Shroff, ``Utility maximization for communication networks with
  multipath routing,'' \emph{IEEE Transactions on Automatic Control}, vol.~51,
  no.~5, pp. 766--781, May 2006.

\bibitem{Wyner}
A.~D. Wyner, ``{The wire-tap channel},'' \emph{Bell Syst. Tech. J.}, vol.~54,
  no.~8, pp. 1355--138, Oct. 1975.

\bibitem{Gopala}
P.~K. Gopala, L.~Lai, and H.~E. Gamal, ``{On the secrecy capacity of fading
  channels},'' \emph{IEEE Trans. Inf. Theory}, vol.~54, no.~10, pp.
  4687--–4698, Oct. 2008.

\bibitem{Liang}
Y.~Liang, H.~Poor, and S.~Shamai, ``{Secure communication over fading
  channels},'' \emph{IEEE Trans. Inf. Theory}, vol.~54, no.~6, pp. 2470–--2492,
  June 2008.

\bibitem{Gungor:infocom:10}
O.~Gungor, J.~Tan, C.~E. Koksal, H.~E. Gamal, and N.~B. Shroff, ``Joint power
  and secret key queue management for delay limited secure communication,'' in
  \emph{Proc. IEEE Conf. Computer Communications (Infocom)}, San Diego, CA,
  March 2010.

\bibitem{Khisti}
A.~Khisti and G.~W. Wornel, ``Secure transmissions with multiple antennas: The
  misome wiretap channel,'' \emph{IEEE Trans. Inf. Theory}, vol.~56, no.~7, pp.
  3088--3014, July 2010.

\bibitem{Shaffiee}
S.~Shaffiee, N.~Liu, and S.~Ulukus, ``Towards the secrecy capacity of gaussian
  mimo wire-tap channel: The 2-2-1 channel,'' \emph{IEEE Trans. Inf. Theory},
  vol.~55, no.~9, pp. 4033--4039, Sept. 2009.

\bibitem{Dong}
L.~Dong, Z.~Han, A.~P. Petropulu, and H.~V. Poor, ``Improving wireless physical
  layer security via cooperating relays,'' \emph{IEEE Trans. on Signal
  Processing}, vol.~58, no.~3, pp. 4033--4039, March 2010.

\bibitem{Koyluoglu:TIT:12}
O.~O. Koyluoglu, C.~E. Koksal, and H.~E. Gamal, ``On secrecy capacity scaling
  in wireless networks,'' \emph{IEEE Trans. Inform. Theory}, vol.~58, no.~5,
  pp. 3000--3015, May 2012.

\bibitem{Goeckel:infocom:12}
C.~Capar, D.~Goeckel, B.~Liu, and D.~Towsley, ``Secret communication in large
  wireless networks without eavesdropper location information,'' in \emph{Proc.
  IEEE Conf. Computer Communications (Infocom)}, Orlando, FL, March 2012, pp.
  1152--1160.

\bibitem{Shamir:Comm:79}
A.~Shamir, ``{How to share a secret},'' \emph{Commun. ACM}, vol.~22, no.~11,
  pp. 612--613, Nov. 1979.

\bibitem{Lou:Infocom:04}
W.~Lou, W.~Liu, and Y.~Fang, ``Spread: enhancing data confidentiality in mobile
  ad hoc networks,'' in \emph{Proc. IEEE Conf. Computer Communications
  (Infocom)}, Hong Kong, March 2004, pp. 2404--2413.

\bibitem{Cai:ISIT:02}
N.~Cai and R.~Yeung, ``Secure network coding,'' in \emph{Proc. IEEE Int.
  Symposium Inform. Theory}, Lausanne, Switzerland, June 2002.

\bibitem{Feldman}
J.~Feldman, T.~Malkin, R.~Servedio, and C.~Stein, ``On the capacity of secure
  network coding,'' in \emph{Proc. Allerton Conf. on Communication, Control,
  and Computing}, Monticello, IL, Sep. 2004.

\bibitem{T_Cui}
T.~Cui, T.~Ho, and J.~Kliewer, ``On secure network coding with nonuniform or
  restricted wiretap sets,'' \emph{IEEE Trans. Inf. Theory}, vol.~59, no.~1,
  pp. 166--176, Jan. 2013.

\bibitem{Abuzainab}
N.~Abuzainab and A.~Ephremides, ``{Secure Distributed Information Exchange},''
  \emph{IEEE Trans. Inf. Theory}, vol.~60, no.~2, pp. 1126--1135, Feb. 2014.

\bibitem{Peron}
E.~Peron, ``Information-theoretic secrecy for wireless networks,'' Ph.D.
  dissertation, Ecole polytechnique federale de Lausanne(EPFL), Switzerland,
  2009.

\bibitem{Infocom:Peron}
E.~Perron, S.~Diggavi, and E.~Telatar, ``On cooperative wireless network
  secrecy,'' in \emph{Proc. IEEE Conf. Computer Communications (Infocom)},
  vol.~4, Rio de Janeiro, Brazil, Sep. 2009, pp. 1935--1943.

\bibitem{Koksal}
C.~E. Koksal, O.~Ercetin, and Y.~Sarikaya, ``Control of wireless networks with
  secrecy,'' \emph{IEEE/ACM Trans. on Networking}, vol.~21, no.~1, pp.
  324--337, Feb. 2013.

\bibitem{A_Eryilmaz}
A.~Eryilmaz, R.~Srikant, and J.~R. Perkins, ``{Stable scheduling policies for
  fading wireless channels},'' \emph{IEEE Trans. Inf. Theory}, vol.~13, no.~2,
  pp. 411--424, Apr. 2005.

\bibitem{Manikandan}
C.~Manikandan, S.~Bhashyam, and R.~Sundaresan, ``{Cross-layer scheduling with
  infrequent channel and queue measurements},'' \emph{IEEE Trans. on Wireless
  Communications}, vol.~8, no.~12, pp. 5737--5742, Dec. 2009.

\bibitem{Sanghavi}
S.~Sanghavi, D.~Shah, and A.~Willsky, ``{Message-passing for maximum weight
  independent set},'' \emph{IEEE Trans. Inf. Theory}, vol.~55, no.~11, pp.
  4822--4834, Nov. 2009.

\bibitem{Hoepman}
J.~Hoepman, ``{Simple distribute weighted matchings},'' Oct. 2004, available at
  http://arxiv.org /abs/cs/0410047.

\bibitem{El_Gamal_NIT}
A.~E. Gamal and Y.~Kim, \emph{Network Information Theory}.\hskip 1em plus 0.5em
  minus 0.4em\relax Cambridge University Press, 2011.

\end{thebibliography}

\appendices
\vspace{-0.1in}
\section{Proof of Theorem~\ref{lemma1}}
\label{proof:ach_rate}

%\proof
The variant of Wyner encoding strategy we use is based on random coding and binning~\cite{El_Gamal_NIT}. First, let us describe this strategy.
To begin, node $s$ generates $2^{N({R}_s-\delta)}$ random binary
sequences. Then, it assigns each random binary sequence to one of
$2^{NR_s^{\textit{conf}}}$ bins, so that each bin contains exactly
$2^{N({R}_s-R_s^{\textit{conf}}-\delta)}$ binary sequences. We call
the sequences associated with a bin, the {\em randomization
sequences} of that bin. Each bin of source $s$ is one-to-one matched
with a confidential message $W_s \in \{1,\ldots
,2^{NR_s^{\textit{conf}}}\}$ randomly and this selection is revealed
to the destination and all nodes before the communication starts.
Then, the stochastic encoder of node $s$ selects one of the
randomization sequences associated with each bin at random,
independently and uniformly over all randomization sequences
associated with that bin.  Whenever a message is selected by node
$s$, this particular randomization message is used. This selection
is not revealed to any of the nodes nor to the destination.

%We assume that the confidential message is transmitted to the
%destination in $N_s$ blocks. For the proof, we need to keep track
%the information transferred to the destination and information
%accumulated at intermediate nodes in each of those $N_s$ blocks.
Without loss of generality, we assume that the routes between source
and destination have $p$ multi-paths and each path may consist of a
different number of hops. Let us denote $H_k$ as the number of hops
along $k$th path. Let us denote the randomization sequence of
message $W_s$ as $W_s^r$ for source $s$, and denote the transmitted
vector of channel symbols from node located at hop $i$ along path
$k$ as $\vc{X_{i,k}^s}= [X_{i,k}^s(1), \ldots, X_{i,k}^s(N_s)]$,
where $X_{i,k}^s(t)$ represents the transmitted vector of $N_1$
symbols in block $t$ at hop $i$ along path $k$. Note that
transmitted vectors in the first hop only consists of transmission
of the source $s$. The received signal at intermediate {\em relay}
node located at hop $i$ along path $k$ is $\vc{Y_{i,k}^s} =
[Y_{i,k}^s(1), \ldots, Y_{i,k}^s(N_s)]$, where $Y_{i,k}^s(t)$
represents the received vector of symbols at nodes of hop $i$ along
path $k$ in block $t$. %Note that the vector of symbols in the last
%hops, i.e., $\left[\vc{Y_{H_1,1}^s}, \ldots,
%\vc{Y_{H_p,p}^s}\right]$ is received by the destination.
Also, the
received signal at overhearing neighbor node $j$ when a node at hop
$i$ along path $k$ is transmitting, is $\vc{Z_{i,k}^{j,s}} =
[Z_{i,k}^{j,s}(1), \ldots, Z_{i,k}^{j,s}(N_s)]$. For notational
convenience, let us define $\vc{X^{s}} = [\vc{X_{1,1}^{s}}, \ldots,
\vc{X_{H_1,1}^{s}},\ldots, \vc{X_{1,p}^{s}}, \ldots,
\vc{X_{H_p,p}^{s}}]$ and $\vc{Z^{j,s}} = [\vc{Z_{1,1}^{j,s}},
\ldots, \vc{Z_{H_1,1}^{j,s}},\ldots, \vc{Z_{1,p}^{j,s}}, \ldots,
\vc{Z_{H_p,p}^{j,s}}]$ as transmitted vector of symbols  by all
nodes and the total received signal by the overhearing neighbor node
$j$, respectively.
%The equivocation analysis follows directly for a given scheduling/routing decision.

Next, we provide the equivocation analysis for a given joint scheduling and routing decision. Note that, our objective is to find a value for $R_s^{conf}$ for which perfect secrecy is achievable. To achieve perfect secrecy, we require $R_s-R_s^{conf}$ to be lower bounded by the conditional entropy $H(W_s|\vc{Z^{j,s}})$. For any given intermediate node $j$, the following can be written for the conditional entropy:

%the transmitted vector of channel symbols as $\vc{X_s} = [X_s(1),
%\ldots, X_s(N_s)]$, where $X_s(t)$ represents the transmitted vector
%of $N_1$ symbols in block $t$.

%The received signal at intermediate {\em relay} node $j$ is
%$\vc{Y_j^s} = [Y_j^s(1), \ldots, Y_j^s(N_s)]$, where $Y_j^s(t)$
%represents the received vector of symbols at node $j$ in block $t$.
%Also, the received signal at overhearing neighbor node $i$ is
%$\vc{Z_i^s} = [Z_i^s(1), \ldots, Z_i^s(aN_s)]$.  Assume that there
%are $n$ overhearing neighbor nodes of source $s$. The equivocation
%analysis follows directly for a given scheduling/routing decision.
%For any given intermediate node $j$, conditional entropy is given as

\vspace{-0.15in} \small
\begin{align}
&H(W_s|\vc{Z^{j,s}}) = I(W_s;\vc{Y_{1,1}^s}, \ldots, \vc{Y_{1,p}^s} | \vc{Z^{j,s}}) + H(W_s|\vc{Z^{j,s}},\vc{Y_{1,1}^s}, \ldots, \vc{Y_{1,p}^s}) \nonumber \\
%&\geq I(W_s;\vc{Y_{H_1,1}^s}, \ldots, \vc{Y_{H_p,p}^s}|\vc{Z^{j,s}}) \\
%&=I(W_s;\vc{Y_{H_1,1}^s}, \ldots, \vc{Y_{H_p,p}^s},\vc{Y_{1,1}^s}, \ldots, \vc{Y_{1,p}^s}|\vc{Z^{j,s}})\nonumber \\
%\label{chain_rule1}
%&- I(W_s;\vc{Y_{1,1}^s}, \ldots, \vc{Y_{1,p}^s}|\vc{Z^{j,s}},\vc{Y_{H_1,1}^s}, \ldots, \vc{Y_{H_p,p}^s}) \\
%&\geq I(W_s;\vc{Y_{1,1}^s}, \ldots, \vc{Y_{1,p}^s}|\vc{Z^{j,s}})
%\nonumber \\
%&- I(W_s;\vc{Y_{1,1}^s}, \ldots,
%\vc{Y_{1,p}^s}|\vc{Z^{j,s}},\vc{Y_{H_1,1}^s}, \ldots,
%\vc{Y_{H_p,p}^s}) \\
%&= I(W_s;\vc{Y_{1,1}^s}, \ldots, \vc{Y_{1,p}^s}|\vc{Z^{j,s}})-
%H(W_s|\vc{Z^{j,s}},\vc{Y_{H_1,1}^s}, \ldots,
%\vc{Y_{H_p,p}^s}) \nonumber \\
%&+ H (\vc{Y_{1,1}^s}, \ldots,
%\vc{Y_{1,p}^s}|W_s,\vc{Z^{j,s}},\vc{Y_{H_1,1}^s}, \ldots,
%\vc{Y_{H_p,p}^s}) \nonumber \\
%&\geq I(W_s;\vc{Y_{1,1}^s}, \ldots, \vc{Y_{1,p}^s}|\vc{Z^{j,s}})-
%H(W_s|\vc{Z^{j,s}},\vc{Y_{H_1,1}^s}, \ldots,
%\vc{Y_{H_p,p}^s}) \nonumber\\
%\label{fano_inequaility1}
&\geq I(W_s;\vc{Y_{1,1}^s}, \ldots,
\vc{Y_{1,p}^s}|\vc{Z^{j,s}}) \\
\label{chain_rule}
&=I(W_s,W_s^r;\vc{Y_{1,1}^s}, \ldots, \vc{Y_{1,p}^s}|\vc{Z^{j,s}})-I(W_s^r;\vc{Y_{1,1}^s}, \ldots, \vc{Y_{1,p}^s}|\vc{Z^{j,s}},W_s)\\
&=I(W_s,W_s^r;\vc{Y_{1,1}^s}, \ldots, \vc{Y_{1,p}^s}|\vc{Z^{j,s}})-H(W_s^r|\vc{Z^{j,s}},W_s)\nonumber \\
&+H(W_s^r|\vc{Z^{j,s}},\vc{Y_{1,1}^s}, \ldots,
\vc{Y_{1,p}^s},W_s)\nonumber \\
&\geq I(W_s,W_s^r;\vc{Y_{1,1}^s}, \ldots, \vc{Y_{1,p}^s}|\vc{Z^{j,s}})-H(W_s^r|\vc{Z^{j,s}},W_s)  \\
\label{fano_inequaility} &\geq I(W_s,W_s^r;\vc{Y_{1,1}^s}, \ldots,
\vc{Y_{1,p}^s}|\vc{Z^{j,s}}) -N\epsilon_1
\\ \label{chain_rule_markov}
 &=
I(\vc{X^{s}};\vc{Y_{1,1}^s}, \ldots, \vc{Y_{1,p}^s}|\vc{Z^{j,s}})
-I(\vc{X^{s}};\vc{Y_{1,1}^s}, \ldots,
\vc{Y_{1,p}^s}|\vc{Z^{j,s}},W_s,W_s^r)-N\epsilon_1
\\
\label{inequality}
&\geq I(\vc{X^{s}};\vc{Y_{1,1}^s}, \ldots, \vc{Y_{1,p}^s}|\vc{Z^{j,s}})-N(\epsilon_1+\epsilon_2) \\
\label{chain_rule2}
&=I(\vc{X^{s}};\vc{Y_{1,1}^s}, \ldots, \vc{Y_{1,p}^s},\vc{Z^{j,s}})-I(\vc{X^s};\vc{Z^{j,s}})- N(\epsilon_1+\epsilon_2) \\
&\geq I(\vc{X^{s}};\vc{Y_{1,1}^s}, \ldots, \vc{Y_{1,p}^s})-I(\vc{X^s};\vc{Z^{j,s}})- N(\epsilon_1+\epsilon_2) \\
%\label{minimum_cut}
%&\geq \min_{1\leq m\leq H}\{ I(X_m^s;Y_m^s)\}-\sum_{i=1}^H I(X_i^s(t);Z_{i,j}^s(t))- N(\epsilon_1+\epsilon_2) \\
\label{stable_equivalent}
&\geq I(\vc{X^{s}};\vc{Y_{1,1}^s}, \ldots, \vc{Y_{1,p}^s})-\sum_{k=1}^p\sum_{i=1}^{H_k}I(\vc{X_{i,k}^s};\vc{Z_{i,k}^{j,s}})- N(\epsilon_1+\epsilon_2) \\
&\geq\sum_{t=1}^{N_2}\left[ I(\vc{X^s}(t);Y_{1,k}^s(t), \ldots,
Y_{1,p}^s(t)) -
\sum_{k=1}^p \sum_{i=1}^{H_k} I(X_{i,k}^s(t);Z_{i,k}^{j,s}(t))\right] \nonumber \\
\label{iid_process} &-N(\epsilon_1+\epsilon_2) \\
\label{strong_law_large_numbers} &\geq N\left[ \sum_{(s,i) \in L}
\bar{\mu}_{si}^s - \bar{f}_j^s -(\epsilon_1+\epsilon_2)\right]
\end{align}
\normalsize with probability 1, for any positive
$(\epsilon_1,\epsilon_2)$ doublet%and arbitrarily small
%$\delta$
, as $N_1N_s \rightarrow \infty$. %Here,
%\eqref{chain_rule1} is by the chain rule, \eqref{fano_inequaility1}
%follows from the application of Fano's eqality, since we assume that
%the condition C1. holds, which implies that given routing policy is
%devised such that the decoding constraint in
%\eqref{eq:decoding_constraint} is met.
\eqref{chain_rule} is by the chain rule, \eqref{fano_inequaility}
follows from the application of Fano's equality,
\eqref{chain_rule_markov} follows from the chain rule and that
$(W_s,W_s^r) \leftrightarrow (\vc{X^s}) \leftrightarrow
(\vc{Y_{1,1}^{s}}, \ldots, \vc{Y_{H_1,1}^{s}},\ldots,
\vc{Y_{1,p}^{s}}, \ldots, \vc{Y_{H_p,p}^{s}},\vc{Z^{j,s}})$ forms a
Markov chain, \eqref{inequality} holds since
$I(\vc{X^s};\vc{Y_{1,1}^s}, \ldots,
\vc{Y_{1,p}^s}|\vc{Z^{j,s}},W_s,W_s^r)\leq N\epsilon_2$ as the
transmitted symbols sequences $\vc{X^s}$ is determined w.p.1 given
$(\vc{Z^{j,s}},W_s,W_s^r)$ (with the routing history), and
\eqref{chain_rule2} follows from the chain rule.
\eqref{stable_equivalent} follows from the definition of
transmission across $p$ path and multiple hops,
$I(\vc{X_{i,k}^s};\vc{Y_{l,m}^s}) = 0$ when $i \neq l$ or $k \neq
m$. Hence, $ I(\vc{X^s};\vc{Z^{j,s}}) \leq
\sum_{k=1}^p\sum_{i=1}^{H_k} I(\vc{X_{i,k}^s};\vc{Z_{i,k}^{j,s}})$.
\eqref{iid_process} holds since the fading processes are iid for a
strategy that chooses the transmitted packets injected by the source
to be independent in different blocks, and finally
\eqref{strong_law_large_numbers} follows from
ergodicity.\footnote{Since we are interested in an achievability
scheme, it is sufficient to show that the provided rate is
achievable by any particular strategy.} Noting that
\eqref{strong_law_large_numbers} holds for any $j$ and combining it
with (\ref{eq:condition_R_s}), completes the proof.
%\endproof

%\vspace{-0.15in}
\section{Proof of Theorem \ref{thm:optimalcontrol-1}}
\label{proof:optimalcontrol-1}
%\proof

The optimality of
the algorithm can be shown by applying the Lyapunov optimization
theorem~\cite{Georgiadis}. %Before restating this theorem, we define the
%following parameters. Let $\Bv(t)=(B_1(t),\ldots, B_K(t))$ be the backlog process of the queues in a given queuing system, and let our objective be the maximization of
%time average of a scalar valued function $g(\cdot)$ of another
%process $\Rv(t)$ while keeping $\vc{B}(t)$ finite. Also define
%$\Delta(\Bv(t))=\E{L(\Bv(t+1))-L(\Bv(t))|\Bv(t)}$ as the drift of
%some appropriate Lyapunov function $L(\cdot)$.
%\begin{theorem}{(Lyapunov Optimization)~\cite{Neely}}
%\label{thm:lyap} For the scalar valued function $f(\cdot)$, if there
%exists positive constants $V$, $\epsilon$, $H$, such that for all
%blocks $t$ and all unfinished work vector $\Bv(t)$ the Lyapunov
%drift satisfies: \vspace{-0.15in} \begin{equation} \Delta(\Bv(t)) -
%V \E{f(\Rv(t))|\Bv(t)} \leq H - V g^* -\epsilon \sum_{i=1}^{K}
%B_i(k), \label{eq:optcond}
%\end{equation}
%then the time average utility and queue backlog satisfy:
%\vspace{-0.1in}\begin{align}
%\liminf_{t\rightarrow\infty}\frac{1}{t}\sum_{\tau=0}^{t-1}\E{g(\Rv(\tau))} &\geq g^* - \frac{H}{V} \label{eq:LyapOpt1} \\
%\limsup_{t\rightarrow\infty}\frac{1}{t}\sum_{\tau=0}^{t-1}
%\sum_{i=1}^{K}\E{B_i(\tau)}
%&\leq\frac{H+V(\bar{g}-g^*)}{\epsilon},\label{eq:LyapOpt2}
%\end{align}
%where $g^*$ is the optimal value of $\E{g(\cdot)}$ and $\bar{g} =
%\limsup_{t\rightarrow\infty}\frac{1}{t}\sum_{\tau=0}^{t-1}\E{g(\Rv(\tau))}$.
%\end{theorem}
%Note that we have stated a modified version of the Theorem in
%\cite{Neely}, since we are interested in the average of the utility
%function. The proof is the same excluding the last step given in the
%proof in \cite{Neely}.
We consider queue backlog vectors for commodity $s$ as
$\Qv(t)=(Q_1(t),\ldots, Q_K(t))$, $\Qv^s(t)=(Q^s_1(t),\ldots,
Q^s_n(t))$, and $\Zv^s(t) = (Z^s_1(t),\ldots, Z^s_n(t))$, where $n$
is the number of intermediate relay nodes in the network. Let
$L(\Qv(t),\Qv^s(t),\Zv^s(t))$ be a quadratic Lyapunov function of
real and virtual queue backlogs for commodity $s$ defined as:

\vspace{-0.2in} \small
\begin{equation} L(\mathbf{Q(t)},\mathbf{Q^s(t)},\mathbf{Z^s(t)}) =
\frac{1}{2}\sum_{s\in S} Q_s(t) + \sum_{i=1}^n
\left[(Q^s_i(t))^2+(Z^s_i(t))^2\right] .
\label{eq:lyapunov-function}
\end{equation}
\normalsize

Also consider the one-step expected Lyapunov drift, $\Delta(t)$ for
the Lyapunov function as: \vspace{-0.1in}
\begin{multline} \Delta(t) =
\mathbb{E}\left[ L(\mathbf{Q(t+1)},\mathbf{Q^s(t+1)},\mathbf{Z^s(t+1)}) \right. \\
- \left.
L(\mathbf{\mathbf{Q(t)},Q^s(t)},\mathbf{Z^s(t)})|\mathbf{Q(t)},\mathbf{Q^s(t)},\mathbf{Z^s(t)}\right].
\label{eq:lyapunov-drift}
\end{multline}

The following lemma provides an upper bound on $\Delta(t)$.
\begin{lemma}
\label{lemma:drift-1}

\small
\begin{align}
&\Delta(t)\leq B - \sum_{s\in S} \E{Q_s(t)\left. \left(A_s(t) -
\sum_{\{i|(s,i)\in L\}}
    \mu_{si}^s(t) \right)\ \right| \ Q_s(t)} \nonumber  \\
&- \sum_{s \in S}\sum_{i \notin S,D} \E{Q_i^s(t) \left.
\left(\sum_{j|(i,j)\in L}\mu_{ij}^s(t)- \sum_{i|(i,j)\in L}
\mu_{ji}^s(t)\right)\ \right|\ Q^s_i(t)} \nonumber
\\
&-\sum_{s \in S}\sum_{i \in E}\E{Z^s_i(t)\left.
\left((1-\alpha_s)A_s(t)+\sum_{j \neq i}f_i^{s,j}(t)\right)\
\right|\ Z^s_i(t)}, \label{eq:delta}
\end{align}
\normalsize where $B>0$ is a constant.
\end{lemma}

\proof Since the maximum transmission power is finite, in any
interference-limited system transmission rates are bounded. Let
$\mu_{ij}^{s,\text{max}}$ be the maximum rate over link $(i,j)$ for
commodity $s$, which depends on the channel states. Also assume that
the arrival rate is bounded, i.e., $A_s^{\text{max}}$ is the maximum
number of bits that may arrive in a block for each source. By simple
algebraic manipulations one can obtain a bound for the difference
$\left(Q_s(t+1)\right)^2-\left(Q_s(t)\right)^2$ and also for other
queues to obtain the result in \eqref{eq:delta}.

%Hence, the following inequalities can be obtained for each user
%queue:

%\endproof

Applying the above lemma, we can complete our proof. In particular,
Lyapunov Optimization Theorem \cite{Georgiadis} suggests that a good
control strategy is the one that minimizes the following:
\vspace{-0.1in}
\begin{equation} \Delta^U(t)=\Delta(t) - H \E{\sum_s
\left(U_s(t)\right) |
\left(\mathbf{Q(t)},\mathbf{Q^s(t)},\mathbf{Z^s(t)}\right)} .
\label{eq:deltawithreward}
\end{equation}
By using~\eqref{eq:delta} in the lemma, we obtain an upper bound for
\eqref{eq:deltawithreward}, as follows:

\vspace{-0.2in}\small
\begin{align}
\Delta^U(k) & <  B - \sum_{s \in S} \E{Q_s(t) \left. \left(A_s(t) -
\sum_{\{i|(s,i)\in L\}}
    \mu_{si}^s(t)  \right)\ \right|\ Q^s_i(t)} \nonumber  \\
&- \sum_{s \in S} \sum_{i \notin S,D} \E{Q_i^s(t)\left.
\left(\sum_{j|(i,j)\in L}\mu_{ij}^s(t) - \sum_{i|(i,j)\in
L}\mu_{ji}^s(t)\right)\ \right|\ Q^s_i(t)} \nonumber
\\
&-\sum_{s \in S} \sum_{i \in
E}\E{Z^s_i(t)\left.((1-\alpha_s)A_s(t)+\sum_{j \neq i}
f_i^{s,j}(t))\ \right|\
Z^s_i(t)} \nonumber\\
& -H \E{\sum_s U_s((1-\alpha_s)A_s(t))} \label{drift_final}
\end{align}
\normalsize
%Since $ Q_i^s(t)\left(\sum_{e \in E} \sum_{i:(i,j) \in e}
%\mu_{ij}(t) - \sum_{e \in E} \sum_{i:(j,i) \in e} \mu_{ji}(t)\right)
%= \sum_{e \in E} \sum_{i:(i,j) \in e} \mu_{ij}(t) (Q_i^s(t)-
%Q_j^s(t))$,
It is easy to observe that our proposed dynamic network control
algorithm minimizes the right hand side of \eqref{drift_final} by
rearranging the terms in \eqref{drift_final}.

If the arrival rates and the given encoding rate, $\alpha_s$, are in
the feasible region, it has been shown in \cite{Georgiadis} that
there must exist a stationary scheduling and rate control policy
that chooses the users and their transmission rates independent of
queue backlogs and only with respect to the channel statistics.  In
particular, the optimal stationary policy can be found as the
solution of a deterministic policy if the channel statistics are
known a priori.

Let $U^*$ be the optimal value of the objective function of the
problem (\ref{objective_func_fixed}-\ref{secrecy_const_fixed})
obtained by the aforementioned stationary policy. Also let
${\lambda_s}^*$ be optimal traffic arrival rates found as the
solution of the same problem. In particular, the optimal input rate
${\lambda_s}^*$ could in principle be achieved by the simple
backlog-independent admission control algorithm of new arrival
$A_s(t)$ for a given commodity $s$ in block $t$ independently with
probability $\zeta_s={\lambda_s}^*/\lambda_s$.

Also, since ${\lambda_s}^*$ is in the achievable rate region, i.e.,
arrival rates are strictly interior of the rate region, there must
exist a stationary scheduling and rate allocation policy that is
independent of queue backlogs and satisfies the following:

\vspace{-0.2in}\small
\begin{align}
\sum_{\{i|(s,i)\in L\}}
    \bar{\mu}_{si}^s &\geq {\lambda_s}^*+\epsilon_1 \label{eq:opt-conds-1}
\\ \sum_{j|(i,j)\in L}\bar{\mu}_{ij}^s &\geq
\sum_{i|(i,j)\in L}\bar{\mu}_{ji}^s+\epsilon_2 \label{eq:opt-conds-2}\\
\bar{f}_i^s & \leq (1-\alpha_s){\lambda_s}^*+\epsilon_3 .
\label{eq:opt-conds-3}
\end{align}
\normalsize

Note that as we consider stationary and ergodic policies, long-term
averages in \eqref{eq:opt-conds-1}-\eqref{eq:opt-conds-3} correspond
to expectations of the same variables as in \eqref{drift_final}.
Clearly, any stationary policy should satisfy \eqref{drift_final}.
Recall that our proposed policy minimizes the right hand side (RHS)
of \eqref{drift_final}, and hence, any other stationary policy
(including the optimal policy) has a higher RHS value than the one
attained by our policy. In particular, the stationary policy that
satisfies \eqref{eq:opt-conds-1}-\eqref{eq:opt-conds-3}, and
implements aforementioned probabilistic admission control can be
used to obtain an upper bound for the RHS of our proposed policy.
Inserting \eqref{eq:opt-conds-1}-\eqref{eq:opt-conds-3} into
\eqref{drift_final}, we obtain the following upper bound for our
policy:

\vspace{-0.15in}\small
\begin{align}
RHS<B&-\sum_{s\in S}\epsilon_1\mathbb{E}[Q_s(t)]-\sum_{s\in
S}\sum_{i\notin S,D}\epsilon_2\mathbb{E}[Q_i^s(t)] \nonumber \\
&-\sum_{s\in S}\sum_{i\in E}\epsilon_3\mathbb{E}[Z_i^s(t)]-HU^*.
\nonumber
\end{align}
\normalsize

This is exactly in the form of Lyapunov Optimization Theorem given
in~\cite{Georgiadis}, and hence, we can obtain bounds on the
performance of the proposed policy and the sizes of queue backlogs
as given in Theorem \ref{thm:optimalcontrol-1}.
%\endproof

%
%\section{Proof of Lemma \ref{lemma:opt-stat}}
%\label{proof:opt-stat} \proof It is shown in \cite{Neely:TIT} that
%the optimality is achieved within the class of stationary policies
%$S$, for a large class of network flow problems including fairness
%problems. Since the channel states are chosen from a finite set and
%the set $\{(i,j,\rho)\ |\ i \in \{1,2,\cdots,n\}, j \in
%\{1,2,\cdots,n\}, \alpha \in [0,R^{o,max}]\}$ is closed and bounded, Lemma
%can be proved using similar arguments as in \cite{Neely:TIT}.
%%Let the
%%feasibility region of \eqref{eq:problem} be $\Lambda$ and let
%%{\boldmath $\epsilon$} $ \triangleq (\epsilon\ \epsilon\cdots\
%%\epsilon)$.
%\endproof

\section{Proof of Theorem \ref{thm:optimalcontrol-2}}
\label{proof:optimalcontrol-2}

For the proof, we follow the similar approach used in Appendix
\ref{proof:optimalcontrol-1}, i.e., applying the Lyapunov
optimization theorem. However, here, we use K-slot Lyapunov drift
instead of the one-step expected Lyapunov drift. We again use
quadratic Lyapunov function as in \eqref{eq:lyapunov-function}.
Assume that nodes transmit its exact queue values in every $K$
slots, i.e., in $\ldots,t-K,t,t+K,\ldots$. Thus, $\hat{Q}(t) =
\hat{Q}(t+\tau)$ for $0\leq \tau \leq K$. Consider the K-step
expected Lyapunov drift, $\hat{\Delta}_K(t)$ as:

\vspace{-0.2in} \small
\begin{multline} \hat{\Delta}_K(t)
=
\mathbb{E}\left[ L(\mathbf{\hat{Q}(t+K)},\mathbf{\hat{Q}^s(t+K)},\mathbf{\hat{Z}^s(t+K)}) \right. \\
- \left.
L(\mathbf{\mathbf{\hat{Q}(t)},\hat{Q}^s(t)},\mathbf{\hat{Z}^s(t)})|\mathbf{\hat{Q}(t)},\mathbf{\hat{Q}^s(t)},\mathbf{\hat{Z}^s(t)}\right].
\label{eq:lyapunov-drift-1}
\end{multline}
\normalsize

By using the result in Lemma \ref{lemma:drift-1}. We bound K-step
Lyapunov drift as:

\vspace{-0.2in}\footnotesize
\begin{align}
&\hat{\Delta}_K(t)\leq- \sum_{\tau=1}^K \sum_{s\in S}
\E{\hat{Q}_s(t)\left. \left(A_s(t+\tau) - \sum_{\{i|(s,i)\in L\}}
    \mu_{si}^s(t+\tau) \right)\ \right| \ \hat{Q}_s(t)} \nonumber  \\
&- \sum_{\tau=1}^K\sum_{s \in S}\sum_{i \notin S,D}
\E{\hat{Q}_i^s(t) \left. \left(\sum_{j|(i,j)\in
L}\mu_{ij}^s(t+\tau)- \sum_{j|(j,i)\in L} \mu_{ji+\tau}^s(t)\right)\
\right|\ \hat{Q}^s_i(t)} \nonumber
\\
&-\sum_{\tau=1}^K\sum_{s \in S}\sum_{i \in E}\E{\hat{Z}^s_i(t)\left.
\left((1-\alpha_s)A_s(t+\tau)+\sum_{j \neq
i}f_i^{s,j}(t+\tau)\right)\ \right|\ \hat{Z}^s_i(t)} + BK ,
\label{eq:K_delta}
\end{align}

\normalsize Note that  the difference of queue sizes in slot $t$ and
queue sizes in slot $t+\tau$ is bounded by $\tau\left(\max
(A_s^{\text{max}},\mu_{ij}^{s,\text{max}})\right)$, i.e.,
$\hat{Q}_s(t)-Q_s(t+\tau) \leq \tau
\left(\max(A_s^{\text{max}},\mu_{ij}^{s,\text{max}})\right)$. Then
we can rewrite \eqref{eq:K_delta} as:

\vspace{-0.2in} \footnotesize
\begin{align}
&\hat{\Delta}_K(t)\leq - \sum_{\tau=1}^K \sum_{s\in S}
\E{Q_s(t+\tau)\left. \left(A_s(t+\tau) - \sum_{\{i|(s,i)\in L\}}
    \mu_{si}^s(t+\tau) \right)\ \right| \ Q_s(t)} \nonumber  \\
&- \sum_{\tau=1}^K\sum_{s \in S}\sum_{i \notin S,D} \E{Q_i^s(t+\tau)
\left. \left(\sum_{j|(i,j)\in L}\mu_{ij}^s(t+\tau)- \sum_{j|(j,i)\in
L} \mu_{ji+\tau}^s(t)\right)\ \right|\ Q^s_i(t)} \nonumber
\\
&-\sum_{\tau=1}^K\sum_{s \in S}\sum_{i \in E}\E{Z^s_i(t+\tau)\left.
\left((1-\alpha_s)A_s(t+\tau)+\sum_{j\neq
i}f_i^{s,j}(t+\tau)\right)\ \right|\
Z^s_i(t)} \nonumber \\
&+BK + B'K(K-1), \label{eq:K_delta1}
\end{align}
\normalsize where $B'=
\frac{\max((A_s^{max})^2,(\mu_{ij}^{s,max})^2)}{2}$. In
\eqref{eq:K_delta1}, we obtain a bound for the K-step Lyapunov
drift. We define one-step expected Lyapunov drift as:

\vspace{-0.2in} \small
\begin{multline} \hat{\Delta}(t)
=
\mathbb{E}\left[ L(\mathbf{\hat{Q}(t+1)},\mathbf{\hat{Q}^s(t+1)},\mathbf{\hat{Z}^s(t+1)}) \right. \\
- \left.
L(\mathbf{\mathbf{\hat{Q}(t)},\hat{Q}^s(t)},\mathbf{\hat{Z}^s(t)})|\mathbf{\hat{Q}(t)},\mathbf{\hat{Q}^s(t)},\mathbf{\hat{Z}^s(t)}\right].
\label{eq:lyapunov-drift-2}
\end{multline}
\normalsize

Then, we obtain one-step Lyapunov drift, $\hat{\Delta}(t)$ from
\eqref{eq:K_delta1} as:

\vspace{-0.2in}\small
\begin{align}
\hat{\Delta}(t)&\leq B + B'(K-1)- \sum_{s\in S} \E{Q_s(t)\left.
\left(A_s(t) - \sum_{\{i|(s,i)\in L\}}
    \mu_{si}^s(t) \right)\ \right| \ Q_s(t)} \nonumber  \\
&- \sum_{s \in S}\sum_{i \notin S,D} \E{Q_i^s(t) \left.
\left(\sum_{j|(i,j)\in L}\mu_{ij}^s(t)- \sum_{j|(j,i)\in L}
\mu_{ji}^s(t)\right)\ \right|\ Q^s_i(t)} \nonumber
\\
&-\sum_{s \in S}\sum_{i \in E}\E{Z^s_i(t)\left.
\left((1-\alpha_s)A_s(t)+\sum_{j\neq i}f_i^{s,j}(t)\right)\ \right|\
Z^s_i(t)}
\nonumber \\
&= B'(K-1) + \mbox{RHS  of (26)} \label{eq:K_delta2}
\end{align}

\normalsize After obtaining bound on the $\hat{\Delta}(t)$, we can
obtain bounds on the performance of the proposed policy and the
sizes of queue backlogs as given in Theorem
\ref{thm:optimalcontrol-2} by following the same lines in Appendix
\ref{proof:optimalcontrol-1}.
%\endproof
%$\Delta_K(t)$.
%\begin{lemma}
%\label{lemma:drift-2}

%\
%\normalsize where $B>0$ and $B'>0$ is a constant.
%\end{lemma}

%\proof The proof of Lemma \ref{lemma:drift-2} follows that the
%difference of queue sizes in slot $t$ and queue sizes in slot $t+k$
%is bounded by $k\max{A_s^{\text{max}},\mu_{ij}^{s,\text{max}}}$. By
%writing $Q$

%Since the maximum transmission power is finite, in any
%interference-limited system transmission rates are bounded. Let
%$\mu_{ij}^{s,\text{max}}$ be the maximum rate over link $(i,j)$ for
%commodity $s$, which depends on the channel states. Also assume that
%the arrival rate is bounded, i.e., $A_s^{\text{max}}$ is the maximum
%number of bits that may arrive in a block for each source. By simple
%algebraic manipulation one can obtain a bound for the difference
%$\left(Q_s(t+1)\right)^2-\left(Q_s(t)\right)^2$ and also for other
%queues to obtain the result in \eqref{eq:delta}

%Hence, the following inequalities can be obtained for each user
%queue:

\vspace{-0.7in}
\begin{IEEEbiography}[{\includegraphics[width=1in,height=1.25in,clip,keepaspectratio]{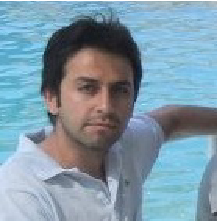}}]{Yunus
Sarikaya} received the BS, MS and PhD degrees in telecommunications
engineering from Sabanci University, Istanbul, Turkey, in 2006, 2008
and 2014, respectively.

Since 2015, he has been with the Electrical and Computer
Engineering Department at Monash University, VIC, as an Senior
Research Associate. He was also a visiting scholar at The Ohio State
University, OH, USA. His research interests include  optimal control of wireless
networks, stochastic optimization, and information theoretical
security.
\end{IEEEbiography}

\vspace{-0.7in}
\begin{IEEEbiography}[{\includegraphics[width=1in,height=2.5in,clip,keepaspectratio]{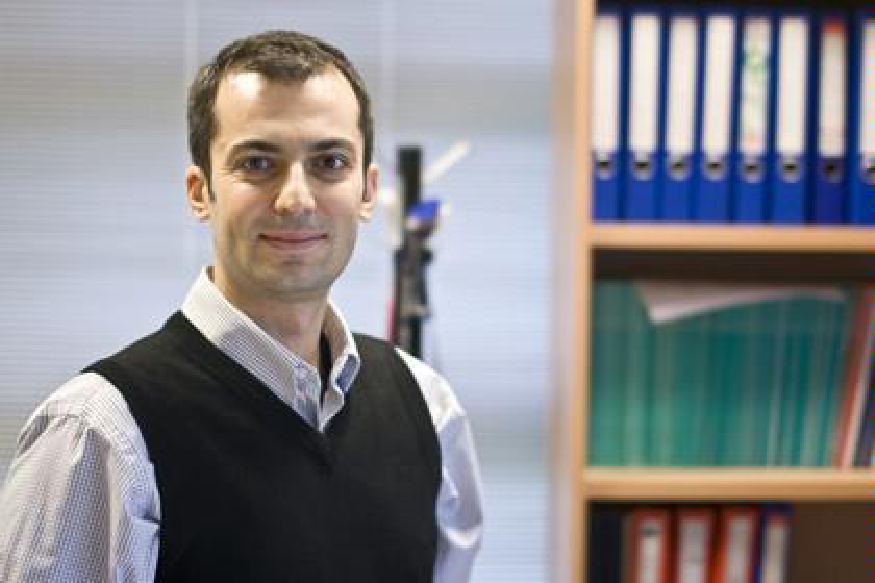}}]{Ozgur
Ercetin} received the BS degree in electrical and electronics
engineering from the Middle East Technical University, Ankara,
Turkey, in 1995 and the MS and PhD degrees in electrical engineering
from the University of Maryland, College Park, in 1998 and 2002,
respectively. Since 2002, he has been with the Faculty of
Engineering and Natural Sciences, Sabanci University, Istanbul. He
was also a visiting researcher at HRL Labs, Malibu, CA, Docomo USA
Labs, CA, and The Ohio State University, OH.  His research interests
are in the field of computer and communication networks with
emphasis on fundamental mathematical models, architectures and
protocols of wireless systems, and stochastic optimization.
\end{IEEEbiography}

\vspace{-0.6in}
\begin{IEEEbiography}[{\includegraphics[width=1in,height=1.25in,clip,keepaspectratio]{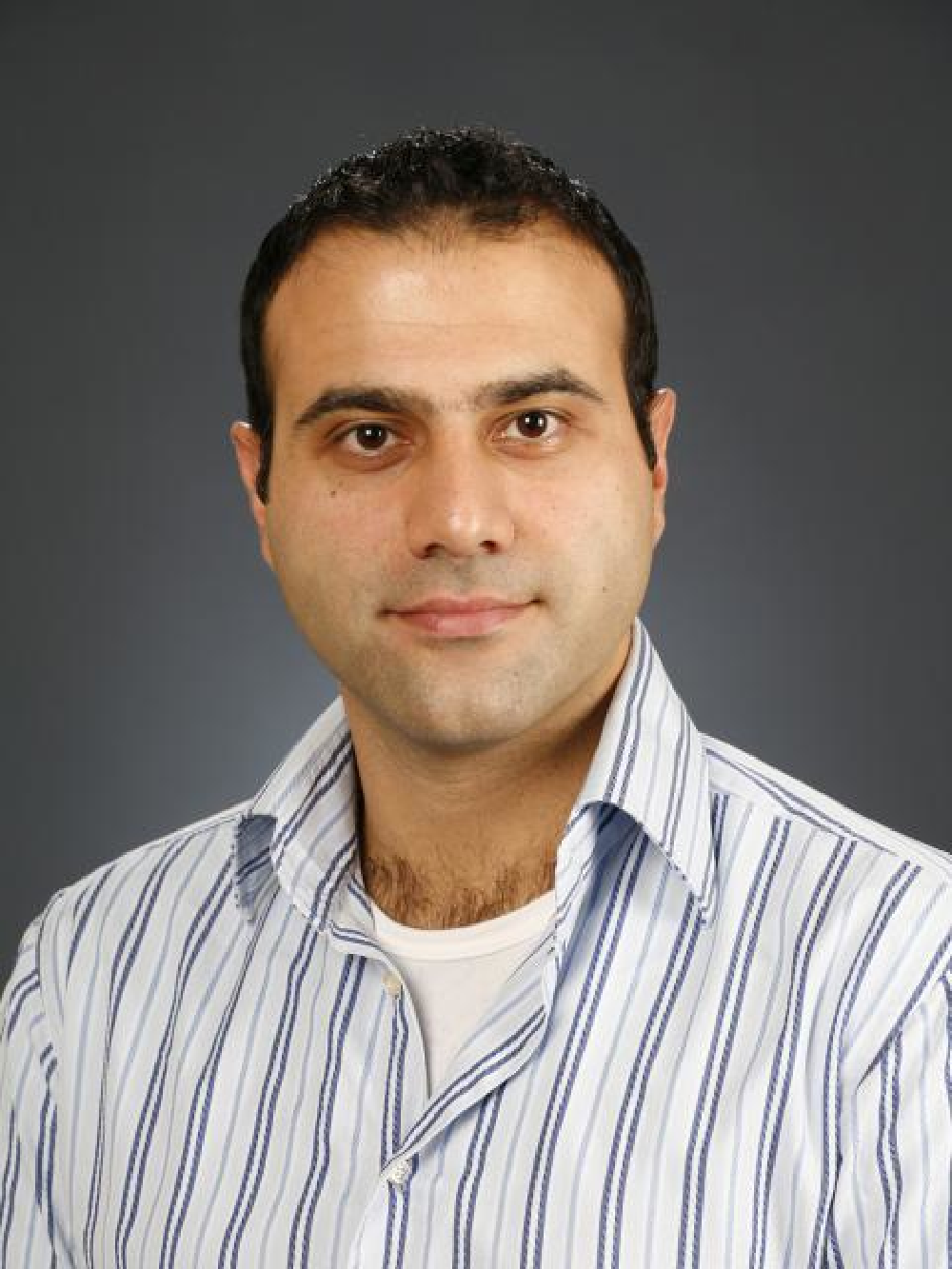}}]{C.~Emre Koksal}
Can Emre Koksal (S'96-M'03-SM'13) received the B.S. degree in
Electrical Engineering from the Middle East Technical University in
1996, and the S.M. and Ph.D. degrees from MIT in 1998 and 2002,
respectively, in Electrical Engineering and Computer Science. He was
a Postdoctoral Fellow at MIT until 2004, and a Senior Researcher at
EPFL until 2006. Since then, he has been with the Electrical and
Computer Engineering Department at Ohio State University, currently
as an Associate Professor. His general areas of interest are
wireless communication, communication networks, information theory,
stochastic processes, and financial economics.

He is the recipient of the National Science Foundation CAREER Award
in 2011, the OSU College of Engineering Lumley Research Award in
2011, the co-recipient of an HP Labs - Innovation Research Award in
2011, and a finalist in Bell Labs Innovation Prize in 2014. The
paper he co-authored was a best student paper candidate in MOBICOM
2005. Currently, he is an Associate Editor for IEEE Transactions on
Information Theory, IEEE Transactions on Wireless Communications,
and Elsevier Computer Networks.
\end{IEEEbiography}

%\newpage
%\input{appendix_opt2}
%\input{figures}

\end{document}